\documentclass[12pt]{article}
\def\theequation{\thesection.\arabic{equation}}
\usepackage{latexsym,here}
\usepackage{graphicx}
\usepackage{epsfig}
\newcommand{\be}{\begin{equation}}
\newcommand{\ee}{\end{equation}}
\newcommand{\bear}{\begin{eqnarray}}
\newcommand{\ear}{\end{eqnarray}}

\def\vec#1{\mathchoice{\mbox{\boldmath$\displaystyle\bf#1$}}
{\mbox{\boldmath$\textstyle\bf#1$}}
{\mbox{\boldmath$\scriptstyle\bf#1$}}
{\mbox{\boldmath$\scriptscriptstyle\bf#1$}}}

\begin{document}

\bibliographystyle{unsrt}

\begin{flushright}
HD-THEP-2001-44
\end{flushright}
\vspace{2cm}
\begin{center}
{\Large 
Effective field theory approach}\\
\vspace{.3cm}
{\Large to structure functions at small $x_{Bj}$}
\end{center}
\vspace{1cm}
\normalsize
\begin{center}
O. Nachtmann\footnote{e-mail: O.Nachtmann@thphys.uni-heidelberg.de}\\
\vspace{0.5cm}
Institut f\"ur Theoretische Physik\\ 
Universit\"at Heidelberg\\
Philosophenweg 16\\
D-69120 Heidelberg, Germany\\
\vspace{0.5cm}
October 2001
\end{center}
\vspace{1cm}
\bigskip

\begin{abstract}
We relate the structure functions of deep inelastic lepton-nucleon
scattering to current-current correlation functions in a Euclidean
field theory depending on a parameter $r$. The $r$-dependent
Hamiltonian of the theory is $P^0-(1-r)P^3$, with $P^0$ the usual
Hamiltonian and $P^3$ the third component of the momentum operator. We
show that small $x_{Bj}$ in the structure functions correspond to the
small $r$ limit of the effective theory. We argue that for $r\to 0$
there is a critical regime of the theory  where simple scaling
relations should hold. We show that in this framework Regge behaviour
of the structure functions obtained with the hard pomeron ansatz
corresponds to a scaling behaviour of the matrix elements in the 
effective
theory where the intercept of the hard pomeron appears  as a critical
index.
Explicit expressions for various analytic continuations of structure
functions and matrix elements are given as well as path integral
representations for the matrix elements in the effective theory.
Our aim is to  provide a framework for truely nonperturbative
calculations of structure functions at  small $x_{Bj}$ for arbitrary $Q^2$.
\end{abstract}

\newpage
\section{Introduction}
In this article we shall discuss the small $x_{Bj}$ behaviour of the structure
functions of deep inelastic lepton-nucleon scattering (DIS). The
findings of 
the experiments H1 and ZEUS at HERA (for recent summaries see
\cite{1}-\cite{3}) 
have brought this topic to the forefront of theoretical
interest. Soon more data will come from HERA2. There are numerous
suggestions for the theoretical description of the small $x_{Bj}$ behaviour
of the structure functions, see for instance \cite{301, 302, 303}.
Let us just mention a few of these approaches with representative
references.

As the first group of approaches let us mention the ones based
on perturbative QCD, which allows one to derive evolution equations
for the structure functions. It should be kept in mind that already
in the derivation of these evolution equations one has to make
various assumptions and their practical use involves further 
approximations. 

Most popular and widely used is the DGLAP equation \cite{304}
to calculate the evolution of the structure functions with $Q^2$
(see for instance \cite{305, 1,3}). Improvements of the 
DGLAP method in fixed order in the strong coupling parameter $\alpha_s$,
involving resummations to all orders in $\alpha_s$, have
recently been proposed \cite{306}. Another time-honoured approach
is based on the BFKL equation \cite{307} which is described in
detail for instance in \cite{308}. However, very large higher order
corrections have been found in this approach \cite{308a}. Different
recipes for dealing with this problem have been proposed \cite{308b}.

Other approaches make more assumptions and could be called QCD-based
or -in\-spired models. Very popular at the moment are dipole models
\cite{309,312}. Other approaches are based on the semiclassical
approximation \cite{310,303,311,313} and on the colour glass
condensate idea \cite{314,315}.

Quite a different type of approach is based on Regge theory. It was 
shown in \cite{316,317} that the small $x_{Bj}$ behaviour of the
structure functions can be well described using two pomerons, a hard
and a soft one. Both pomerons are assumed to behave like simple
Regge poles with linear trajectories. 

In this article we continue the investigations of 
the approach \cite{12} where the behaviour of the structure 
functions at small $x_{Bj}$ is related to that of matrix elements
in an effective Euclidean field theory. In \cite{12} this was 
explored for a model scalar field theory and it was argued that the 
limit $x_{Bj}\to 0$ corresponds to critical behaviour in the effective theory.
Here we extend these considerations to the case of QCD. The aim of this
approach is to provide a framework where the small $x_{Bj}$ behaviour
of the structure functions can be calculated from first principles
using truely nonperturbative methods, for instance lattice methods.

Our article is organised as follows. In section 2 we discuss kinematics,
the reduced matrix elements free of kinematical singularities and the
analytic continuation from the real to the imaginary $\nu$-axis.
Section 3 deals with the Deser-Gilbert-Sudarshan (DGS) representation
which we use for further analytic continuations. In section 4 we discuss
our effective Hamiltonians and Lagrangians for QCD, both in Minkowski and Euclidean space. Section 5 contains phenomenological applications
and section 6 our conclusions.

\section{Kinematics and analytic continuation in the $\nu$-plane}

In this section we recall first some  definitions and results
from DIS. The central object of our study is the forward virtual Compton
scattering amplitude for nucleons. The absorptive part of this
amplitude gives for real $\nu$  the measurable cross sections of DIS. Our
theoretical investigations will concentrate on the amplitude at imaginary 
values of $\nu$, from which nevertheless we can obtain information on these 
cross sections. 

\subsection{Kinematics}

We study the forward virtual Compton scattering amplitude (Fig. 1)
\be\label{2.1}
\gamma^*(q)+N(p)\to \gamma^*(q)+N(p),\ee
where $N$ stands for proton or neutron. We consider only  
spacelike virtual photons, $q^2=-Q^2<0$ and the amplitude 
averaged over the nucleon spin. The familiar Feynman amplitude
for (\ref{2.1}) is
\be\label{2.2}
T^{\rm F}_{\mu\nu}(p,q)=\frac{i}{2\pi M}\int d^4x\ e^{iqx}
{\textstyle\frac{1}{2}}\sum_{spins} 
\langle N(p)|{\rm T}^*J_\mu(x)J_\nu(0)|N(p)\rangle.\ee
Here $M$ is the nucleon mass,
$J_\mu(x)$ is the hadronic part  of
the electromagnetic current, and ${\rm T}^*$ indicates the covariant version
of the T product (see for instance \cite{13}). It is understood that only 
the connected part of the matrix element is  taken. All our 
conventions on kinematics
follow \cite{14}.

In the following we shall, however, not work with (\ref{2.2})
but with the retarded amplitude
\bear\label{2.3}
T^{\rm ret}_{\mu\nu}(p,q)=\frac{i}{2\pi M}\int d^4 x\ e^{iqx}
{\textstyle\frac{1}{2}}\sum_{spins}
\langle N(p)|\theta(x^0)[J_\mu(x),J_\nu(0)]_{cov}
|N(p)\rangle,\ear
where we define
\be\label{2.4}
\theta(x^0)[J_\mu(x),J_\nu(0)]_{cov}={\rm T}^*(J_\mu(x)J_\nu(0))-J_\nu
(0)J_\mu(x).\ee
The standard $W_{\mu\nu}$ tensor of DIS is 
\be\label{2.5a}
W_{\mu\nu}(p,q)=\frac{1}{4\pi M}\int d^4 xe^{iqx} {\cal M}_{\mu\nu}(x,p),\ee
where
\be\label{2.5b}
{\cal M}_{\mu\nu}(x,p)={\textstyle\frac{1}{2}}\sum_{spins} \langle N(p)|
J_\mu(x)J_\nu(0)|N(p)\rangle.\ee
The expansion of $W_{\mu\nu}$ in terms of the structure functions
$W_{1,2}$ reads for $pq>0$:
\bear\label{2.6}
&&W_{\mu\nu}(p,q)=\left(-g_{\mu\nu}+\frac{q_\mu q_\nu}{q^2}\right)
W_1(\nu,Q^2)\nonumber\\
&&+\frac{1}{M^2}\left(p_\mu-\frac{(pq)q_\mu}{q^2}\right)
\left(p_\nu-\frac{(pq)q_\nu}{q^2}\right)W_2(\nu,Q^2),\nonumber\\
&&\nu=pq/M,\quad Q^2=-q^2.\ear
In the usual way we define the structure functions $F_{1,2,L}$
as
\bear\label{2.7a}
F_1(x_{Bj}, Q^2)&=& 2MW_1(\nu,Q^2),\nonumber\\
F_2(x_{Bj},Q^2)&=&\nu W_2(\nu,Q^2),\nonumber\\
F_L(x_{Bj},Q^2)&=&F_2(x_{Bj},Q^2)
 -(1+\frac{4M^2}{Q^2}x^2_{Bj})^{-1}x_{Bj}F_1(x_{Bj},Q^2),\nonumber\\
x_{Bj}&=&\frac{Q^2}{2M\nu},\ear
and the transverse and longitudinal cross sections
\bear\label{2.7b}
\frac{K}{4\pi^2\alpha}\sigma_T(\nu,Q^2)&=&W_1(\nu,Q^2)\nonumber\\
&=&\frac{1}{2M}F_1(x_{Bj},Q^2),\nonumber\\
\frac{K}{4\pi^2\alpha}\sigma_L(\nu,Q^2)&=&\frac{\nu^2+Q^2}{Q^2}
W_2(\nu,Q^2)-W_1(\nu,Q^2)\nonumber\\
&=&\frac{1}{2Mx_{Bj}}(1+\frac{4M^2}{Q^2}x^2_{Bj})F_L(x_{Bj},Q^2),
\nonumber\\
K&=&\nu-\frac{Q^2}{2M}.\ear
The expansions for $T_{\mu\nu}^{\rm F}$ (\ref{2.2}) and 
$T^{\rm ret}_{\mu\nu}$ (\ref{2.3}) read
\bear\label{2.7}
&&T_{\mu\nu}^{\rm F,ret}(p,q)=\left(-g_{\mu\nu}+
\frac{q_\mu q_\nu}{q^2}\right)
T_1^{\rm F,ret}(\nu,Q^2)\nonumber\\
&&+\frac{1}{M^2}\left(p_\mu-\frac{(pq)q_\mu}{q^2}\right)
\left(p_\nu-\frac{(pq)q_\nu}{q^2}\right)T_2^{\rm F,ret}(\nu,Q^2),\ear
where now $-\infty<pq<\infty$. For $Q^2$ fixed, $T_j^{\rm F}(\nu,Q^2)$ and 
$T_j^{\rm ret}(\nu,Q^2)$ $(j=1,2)$, are limits of functions $T_j(\nu,Q^2)$ 
analytic in the cut $\nu$-plane as shown in Fig. 2. The position
of the nucleon poles is $\nu=\pm Q^2/(2M)$, the cuts start at
\be\label{2.8b}
\nu=\pm\nu_t=\pm[Q^2+(M+m_\pi)^2-M^2]/(2M),\ee
where $m_\pi$ is the pion mass.
We have for real $\nu$ and $j=1,2$
\be\label{2.8}
T^{\rm F}_j(\nu,Q^2)=\lim_{\epsilon\to +0}T_j(\nu(1+i\epsilon),Q^2),\ee
\be\label{2.9}
T_j^{\rm ret}(\nu,Q^2)=\lim_{\epsilon\to+0}T_j(\nu
+i\epsilon,Q^2),\ee
\be\label{2.10}
{\rm Im} \, T_j^{\rm F}(\nu,Q^2)=\theta(\nu)W_j(\nu,Q^2)+\theta(-\nu)W_j(-\nu,Q^2),
\ee
\be\label{2.11}
{\rm Im} \, T^{\rm ret}_j(\nu,Q^2)=\theta(\nu)W_j(\nu,Q^2)-\theta(-\nu)
W_j(-\nu,Q^2).\ee

For our purpose it is convenient to define in addition the following
scalar amplitudes:
\bear\label{2.12}
T_a^{\rm ret}(\nu,Q^2)&=&-g^{\mu\nu}T^{\rm ret}_{\mu\nu}(p,q)\nonumber\\
&=& 3T_1^{\rm ret}(\nu,Q^2)-\frac{\nu^2+Q^2}{Q^2}
T_2^{\rm ret}(\nu,Q^2),\ear
\bear\label{2.13}
T^{\rm ret}_b(\nu,Q^2)&=&M^{-2}p^\mu p^\nu T_{\mu\nu}^{\rm ret}(p,q)\nonumber\\
&=&-\frac{\nu^2+Q^2}{Q^2}T^{\rm ret}_1(\nu,Q^2)+\left(\frac{\nu^2+Q^2}{Q^2}
\right)^2 T_2^{\rm ret}(\nu,Q^2).\ear
In a similar way we define for real $\nu$ the functions 
$T^{\rm F}_{a,b}$ and 
$W_{a,b}$ and for complex $\nu$
the functions
$T_{a,b}(\nu,Q^2)$. The latter are analytic in the cut $\nu$-plane and have as
boundary values  for ${\rm Im}\ \nu\to+0$ the functions $T_{a,b}
^{\rm ret}(\nu,Q^2)$.

\subsection{Analytic continuation in the $\nu$-plane}

We are interested in the behaviour of $W_{1,2}(\nu,Q^2)$ for fixed $Q^2>0$
and $\nu\to\infty$. 
Instead of investigating the structure functions $W_{1,2}$ directly
we shall study first the behaviour of the functions $T_{1,2}(\nu,Q^2)$ or
equivalently $T_{a,b}(\nu,Q^2)$ for large imaginary $\nu$, that is for
$\nu=i\eta$ with $\eta\to\infty$. Then we use the
Phragm\'en-Lindel\"of theorem
(see theorem 5.64 of \cite{15}) to relate the behaviour of the amplitudes
for large real and imaginary values of $\nu$. This has been discussed
in detail in appendix A of \cite{12}.

We follow now the same strategy as in \cite{12}. We work in the
rest system of the nucleon and choose the direction of $\vec q$ as third
axis of the coordinate system:
\be\label{2.14}
p={M \choose \vec 0}\ , \quad q={\nu \choose {\vec e}_3\sqrt{\nu^2+Q^2}}.
\ee
Using rotational symmetry and the same arguments as in \cite{12}, we can
represent the functions $T_{a,b}(\nu,Q^2)$ for ${\rm Im}\ \nu\geq 0$ 
as follows:
\newpage
\bear\label{2.15}
T_a(\nu,Q^2)&=&-\frac{1}{M\sqrt{\nu^2+Q^2}}\int^\infty_0
dx^0\int^\infty_{-\infty}dx^3x^3
\exp[i x^0\nu-ix^3\sqrt{\nu^2+Q^2}]\nonumber\\
&&\times {\textstyle\frac{1}{2}}
\sum_{spins}\langle N(p)|(-g^{\mu\nu})\theta(x^0)[J_\mu
(x^3\vec e_3,x^0),J_\nu(0)]_{cov}|N(p)\rangle,\\
\label{2.16}
T_b(\nu,Q^2)&=&-\frac{1}{M\sqrt{\nu^2+Q^2}}\int^\infty_0
dx^0\int^\infty_{-\infty}dx^3x^3
\exp[ix^0\nu -ix^3\sqrt{\nu^2+Q^2}]\nonumber\\
&&\times{\textstyle\frac{1}{2}}\sum_{spins}\langle N(p)|
\theta(x^0)[J_0(x^3\vec e_3,x^0),J_0(0)
]_{cov}|N(p)\rangle .\ear
For $\nu=i\eta$ and $\eta>Q$ we get 
\bear\label{2.17}
T_a(i\eta,Q^2)&=&\frac{i}{M\sqrt{\eta^2-Q^2}}\int^\infty_0dx^0
\int^\infty_{-\infty}dx^3x^3
\exp[-x^0\eta +x^3\sqrt{\eta^2-Q^2}]\nonumber\\
&&\times {\textstyle\frac{1}{2}}\sum_{spins}\langle
N(p)|(-g^{\mu\nu})\theta(x^0)
[J_\mu(x^3\vec e_3,x^0),J_\nu(0)]_{cov}|N(p)\rangle,\\
\label{2.18}
T_b(i\eta,Q^2)&=&\frac{i}{M\sqrt{\eta^2-Q^2}}\int^\infty_0dx^0\int^\infty
_{-\infty}dx^3x^3
\exp[-x^0\eta +x^3\sqrt{\eta^2-Q^2}]\nonumber\\
&&\times {\textstyle\frac{1}{2}}\sum_{spins} \langle N(p)|\theta(x^0)
[J_0(x^3\vec e_3,x^0),J_0(0)
]_{cov}|N(p)\rangle .\ear
The commutators in (\ref{2.15})-(\ref{2.18}) vanish for $|x^3|>x^0$
and 
thus the effective integration range is $|x^3|\leq x^0$. 
This makes the integrands in (\ref{2.15}), (\ref{2.16})
exponentially damped for ${\rm Im}\ \nu>0$ and represents
the standard way of analytic continuaton of $T_{a,b}^{{\rm ret}}(\nu,Q^2)$
into the upper half $\nu$-plane. However, the singularities
on the lightcone $|x^3|=x^0$ make it advisable to keep the $x^3$-integrals
in (\ref{2.15})-(\ref{2.18}) as running from $-\infty$ to $+\infty$.
From $T_{a,b}(i\eta,Q^2)$ we can derive $T_{1,2}(i\eta,Q^2)$ using 
(\ref{2.12}), (\ref{2.13}). 

In the next section we will derive another representation for $T_{1,2}
(i\eta,Q^2)$ in which  the integration path avoids the light-cone.

\section{The matrix element ${\cal M}_{\mu\nu}$}
\setcounter{equation}{0}

In this section we list general properties of the matrix element
${\cal M}_{\mu\nu}
(x,p)$ (\ref{2.5b}) and discuss its analyticity properties using the
Deser-Gilbert-Sudarshan (DGS) representation \cite{16,17}. 

\subsection{General properties of ${\cal M}_{\mu\nu}(x,p)$}

Hermiticity of the electromagnetic current gives
\be\label{3.1}
{\cal M}_{\mu\nu}(x,p)^*={\cal M}_{\nu\mu}(-x,p).\ee
Time reversal invariance of QCD requires
\be\label{3.2}
{\cal M}_{\mu\nu}(x,p)={\cal M}^{\nu\mu}(x',p'),\ee
where $x^{'\mu}=x_\mu, p^{'\mu}=p_\mu$.\newline
Current conservation implies
\bear\label{3.3}
\frac{\partial}{\partial x^\mu}{\cal M}^{\mu\nu}(x,p)=0,\nonumber\\
\frac{\partial}{\partial x^\nu}{\cal M}^{\mu\nu}(x,p)=0.\ear
For $x^2<0$ the currents $J_\mu(x)$ and
$J_\nu(0)$ in (\ref{2.5b}) commute, which leads to
\be\label{3.4}
{\cal M}_{\mu\nu}(x,p)={\cal M}_{\nu\mu}(-x,p),\ {\rm for}
\ x^2<0.\ee

The expansion of ${\cal M}_{\mu\nu}(x,p)$ in terms of two causal scalar
functions which are free from kinematical singularities and where
(\ref{3.1}) to (\ref{3.4}) are satisfied automatically \cite{18} reads
\bear\label{3.5}
{\cal M}_{\mu\nu}(x,p)&=&[g_{\mu\nu}\Box-\partial_\mu\partial_\nu]
{\cal M}_1(xp,x^2)\nonumber\\
&&+[p_\mu p_\nu\Box-(p\partial)(p_\mu\partial_\nu+p_\nu\partial_\mu)
+g_{\mu\nu}(p\partial)^2]{\cal M}_2(xp,x^2).\ear
The invariant functions ${\cal M}_j\ (j=1,2)$ satisfy:
\be\label{3.6}
{\cal M}_j(xp,x^2)={\cal M}^*_j(-xp,x^2)\ee
and
\be\label{3.7}
{\cal M}_j(xp,x^2)={\cal M}_j(-xp,x^2)\quad {\rm for}\ x^2<0.\ee

\subsection{Representations of amplitudes in terms of ${\cal M}_{1,2}$}

Inserting (\ref{3.5}) into (\ref{2.5a}) we obtain the structure
functions $W_{1,2}$ of (\ref{2.6}) in terms of the Fourier transforms
of ${\cal M}_{1,2}$:
\bear\label{3.8}
W_1(\nu,Q^2)&=&-\frac{Q^2}{4\pi M}\int d^4x e^{iqx}{\cal M}_1(xp,x^2)
+\frac{(pq)^2}{4\pi M}\int d^4x e^{iqx} {\cal M}_2(xp,x^2),\\
\label{3.9}
W_2(\nu,Q^2)&=&\frac{MQ^2}{4\pi}\int d^4xe^{iqx}{\cal M}_2(xp,x^2).\ear
Note that in (\ref{3.9}) the kinematic zero of $W_2(\nu,Q^2)$ at
$Q^2=0$ is made explicit, as it should be. For the cross sections 
(\ref{2.7b}) we get:
\bear\label{3.9a}
\frac{K}{4\pi^2\alpha}\sigma_T(\nu,Q^2)&=&\frac{1}{4\pi M}\int
d^4xe^{iqx}
 [M^2\nu^2{\cal M}_2(xp,x^2)-Q^2{\cal M}_1
(xp,x^2)],\\
\label{3.9b}
\frac{K}{4\pi^2\alpha}\sigma_L(\nu,Q^2)&=&\frac{Q^2}{4\pi M}\int
d^4xe^{iqx}
 [M^2{\cal M}_2(xp,x^2)+{\cal M}_1(xp,x^2)].\ear
Inserting (\ref{3.5}) into (\ref{2.3}) and using (\ref{2.7}) and
rotational symmetry in the nucleon rest frame, we get
\bear\label{3.10}
T^{\rm ret}_1(\nu,Q^2)&=&-\frac{1}{M\sqrt{\nu^2+Q^2}}\int^\infty_0
dx^0\int^\infty_{-\infty} dx^3 x^3
\exp[i x^0\nu-ix^3\sqrt{\nu^2+Q^2}]\nonumber\\
&&\times \{-Q^2[{\cal M}_1(xp,x^2)-{\cal M}^*_1(xp,x^2)]\nonumber\\
&&+M^2\nu^2[{\cal M}_2(xp,x^2)-{\cal M}_2^*(xp,x^2)]\},\\
\label{3.11}
T^{\rm ret}_2(\nu,Q^2)&=&-\frac{MQ^2}{\sqrt{\nu^2+Q^2}}\int^\infty_0 dx^0\int
^\infty_{-\infty} dx^3x^3
\exp[ix^0\nu -ix^3\sqrt{\nu^2+Q^2}]\nonumber\\
&&\times [{\cal M}_2(xp,x^2)-{\cal M}_2^*(xp,x^2)]\ear
where 
\be\label{3.12}
xp=x^0M\ , \quad x^2=(x^0)^2-(x^3)^2.\ee

In the following we will make a change of variables as in \cite{12}
and replace $(x^0,x^3)$ by $(x^0,r)$, where 
\be\label{3.12a}
r=1-\frac{x^3}{x^0}.\ee
We define
\bear\label{3.12b}
\widetilde{\cal M}^-_j(x^0,r):&=&{\cal M}_j(x^0M,(x^0)^2r(2-r))
\nonumber\\
&&(j=1,2).\ear
With this we get
\bear\label{3.12c}
T_1^{\rm ret}(\nu,Q^2)&=&-\frac{1}{M\sqrt{\nu^2+Q^2}}\int
^\infty_{-\infty}dr(1-r)\nonumber\\
&&\times \int^\infty_0dx^0(x^0)^2\exp[ix^0(\nu-(1-r)\sqrt{\nu^2+Q^2})]
\nonumber\\
&&\times \{-Q^2[\widetilde{\cal M}_1^-(x^0,r)-(\widetilde{\cal
M}_1^-(x^0,r))^*]
\nonumber\\
&&\phantom{\times}+M^2\nu^2[\widetilde{\cal M}_2^-(x^0,r)-(\widetilde
 {\cal M}_2^-
(x^0,r))^*]\},\ear
\bear\label{3.12d}
T_2^{\rm ret}(\nu,Q^2)&=&-\frac{MQ^2}{\sqrt{\nu^2+Q^2}}\int
^\infty_{-\infty}dr(1-r)\nonumber\\
&&\times\int^\infty_0dx^0(x^0)^2\exp[ix^0(\nu-(1-r)\sqrt{\nu^2+Q^2})]
\nonumber\\
&&\times [\widetilde{\cal M}_2^-(x^0,r)-(\widetilde{\cal M}_2^-(x^0,r))^*].
\ear

\subsection{DGS representations for ${\cal M}_{1,2}$}

For the causal functions ${\cal M}_{1,2}(xp,x^2)$ defined in (\ref{3.5})
we can write down DGS representations
\cite{16,17}. The behaviour of the structure functions for small
$x_{Bj}$ as measured in experiments (see chapter 5 below) suggests that
the DGS representation for ${\cal M}_2$ needs no subtraction, whereas
${\cal M}_1$ needs subtractions. Thus we write 
\be\label{3.13}
{\cal M}_2(xp,x^2)=\int^\infty_0ds\int^1_{-1}d\zeta f_2(s,\zeta)
\exp(i\zeta px)\frac{1}{i}\Delta^+(x,s),\ee
where 
\bear\label{3.14}
{\textstyle\frac{1}{i}}\Delta^+(x,s)&=&
\frac{1}{(2\pi)^3}\int d^4k\ e^{-ikx}\theta(k^0)\delta(k^2-s)\nonumber\\
&=&\frac{1}{4\pi^2}\left(\frac{s}{-x^2
+i\epsilon x^0}\right)^{1/2}K_1\left(\sqrt{s(-x^2+i\epsilon x^0)}
\right)\ear
is one of the usual invariant functions given in
terms of the modified Bessel function $K_1$. 
In (\ref{3.13}) $f_2$ is a function for which (\ref{3.6})
and (\ref{3.7}) require
\be\label{3.18a}
f_2(s,\zeta)=f_2(s,-\zeta)=f_2^*(s,\zeta).\ee
Thus, (\ref{3.13}) can also be written as
\be\label{3.18b}
{\cal M}_2(xp,x^2)=\int^\infty_0ds\int^1_{-1}d\zeta f_2(s,\zeta)
\cos(\zeta px)\frac{1}{i}\Delta^+(x,s).\ee
For ${\cal M}_1$ we write down a subtracted DGS representation:
\be\label{3.18c}
{\cal M}_1(xp,x^2)=\int^\infty_0ds\left\{ f_1^{(0)}(s)
+\int^1_{-1}d\zeta f_1(s,\zeta)[\cos(\zeta px)-1]\right\}
\frac{1}{i}\Delta^+(x,s)\ee
where
\bear\label{3.18d}
f_1^{(0)}(s)=\left(f_1^{(0)}(s)\right)^*,\nonumber\\
f_1(s,\zeta)=f_1(s,-\zeta)=f_1^*(s,\zeta).\ear

In \cite{16,17}
the DGS representation is primarily discussed for the matrix
element of the commutator and the T-product of two currents. From
the DGS representation of the T-product we get immediately 
the representation for the ordinary products of currents used here.

Let us now switch to the matrix elements $\widetilde{\cal M}^-_j(x^0,r)$
(\ref{3.12b}) to discuss their behaviour in $r$ for fixed $x^0$.
For $\widetilde{\cal M}_2^-$  we get from (\ref{3.13})-(\ref{3.18b}) for
$x^0>0$:
\bear\label{3.15}
\widetilde{\cal M}_2^-(x^0,r)&\equiv& \widetilde{\cal M}_2(x^0,r,\epsilon)
\nonumber\\
&=&\int^\infty_0ds\int^1_{-1}d\zeta f_2(s,\zeta)\cos(\zeta Mx^0)\nonumber\\
&&\times \frac{1}{4\pi^2}\frac{s^{1/2}}{x^0}\left(-r(2-r)
+i\epsilon\right)^{-1/2}
K_1\left(s^{1/2}x^0(-r(2-r)+i\epsilon)^{1/2}\right).\nonumber\\
&&{}
\ear
For clarity we keep in the following discussion the
$\epsilon$-parameter $(\epsilon>0)$ as an explicit argument. A similar
expression is obtained for $\widetilde{\cal M}_1^-$ from (\ref{3.18c}).

We will now make the assumption that the weight functions $f_j(s,\zeta)$
are sufficiently well behaved such that the representation
(\ref{3.15}) for $\widetilde {\cal M}_2(x^0,r,\epsilon)$ 
and the analogous one for $\widetilde{\cal M}_1$ allow an
analytic continuation in $r$ at fixed $x^0$ and $\epsilon$.
The singularity structure of $\widetilde{\cal M}_j(x^0,r,\epsilon)$ in the
$r$-plane can then be read off from (\ref{3.15}). There are two 
branch points, at $r=1\mp\sqrt{1-i\epsilon}$, giving for $\epsilon\ll1$,
\be\label{3.16}
r\approx \frac{i}{2}\epsilon\ , \quad{\rm and}\quad r\approx 2-\frac{i}
{2}\epsilon.\ee
We draw the associated cuts to the right (see Fig. 3a).
The cut structure of
\be\label{3.17}
(\widetilde{\cal M}_j(x^0,r^*,\epsilon))^*=
\widetilde{\cal M}_j(x^0,r,-\epsilon)\ee
is shown in Fig. 3b. Looking now at (\ref{3.12c}) and (\ref{3.12d})
we see that the $r$-integrations run on the real $r$-axis from
$r=-\infty$ to $r=+\infty$ over $\widetilde{\cal M}_j(x^0,r,\epsilon)$
and from $r=\infty$ back to $r=-\infty$ over $\widetilde{\cal M}_j
(x^0,r,-\epsilon)$.
Deforming the contours slightly we can obtain $T_{1,2}^{\rm ret}$
as integrals 
over the analytic functions
\be\label{3.18}
\widetilde{\cal M}_j(x^0,r):=\widetilde{\cal M}_j(x^0,r,0)\ee
along a curve $C$ (see Fig. 4). The curve $C$ comes from
the second sheet, moves onto the first sheet between
$r=0$ and 2, circles $r=0$ and moves again onto another sheet
between $r=0$ and 2. We get
\bear\label{3.19}
T_1^{\rm ret}(\nu,Q^2)&=&-\frac{1}{M\sqrt{\nu^2+Q^2}}
\int_C dr(1-r)\nonumber\\
&&\times \int^\infty_0 dx^0(x^0)^2\exp[ix^0(\nu-(1-r)\sqrt{\nu^2+Q^2})]
\nonumber\\
&&\times \{-Q^2\widetilde{\cal M}_1(x^0,r)+M^2\nu^2\widetilde{\cal M}_2(x^0,r)
\},\ear
\bear\label{3.20}
T_2^{\rm ret}(\nu,Q^2)&=&-\frac{MQ^2}{\sqrt{\nu^2+Q^2}}\int_C
dr(1-r)\nonumber\\
&&\times \int^\infty_0dx^0(x^0)^2\exp [ix^0(\nu-(1-r)\sqrt{\nu^2+
Q^2})]\  \widetilde{\cal M}_2(x^0,r).\nonumber\\
&&{}\ear
Note that the  matrix elements $\widetilde {\cal M}^-_j(x^0,r)$ of
(\ref{3.12b}) for $0<r<2$ are the limits of the analytic functions
$\widetilde{\cal M}_j(x^0,r)$ below the cut:
\be\label{3.20a}
\widetilde{\cal M}_j^-(x^0,r)=\lim_{\epsilon\to+0}
\widetilde{\cal M}_j(x^0,r-i\epsilon),
\qquad (0<r<2).\ee

We can now use (\ref{3.19}), (\ref{3.20}) to perform the analytic 
continuation in $\nu$ to the upper half $\nu$-plane in an alternative
way to that of section 2.2. We 
move the curve $C$ arbitrarily close to the positive 
real $r$-axis, $r>0$, and 
use the inequalities (A.1) of \cite{12} to show that the
exponentials in (\ref{3.19}) and (\ref{3.20}) are
damped for ${\rm Im}\ \nu>0$. We are interested in $T_{1,2}$ for large
imaginary $\nu$. With $\nu=i\eta$,
$\eta>Q$, the exponentials can be written as
\be\label{3.21}
\exp[ix^0(\nu-(1-r)\sqrt{\nu^2+Q^2})]|_{\nu=i\eta}
=\exp[-x^0/\bar x(\eta,r)],\ee
\be\label{3.22}
\bar x(\eta,r)=[\eta-(1-r)\sqrt{\eta^2-Q^2}]^{-1}=
\frac{1}{\eta}\frac{1+\bar r}{r+\bar r},\ee
\be\label{3.23}
\bar r=\frac{Q^2}{2\eta^2}\frac{2}{1-\frac{Q^2}{\eta^2}+\sqrt{1
-\frac{Q^2}{\eta^2}}}.\ee
With (\ref{3.21}) we get for the analytically continued functions
$T_{1,2}$ from (\ref{3.19}) and (\ref{3.20})
\bear\label{3.24}
T_1(i\eta,Q^2)&=&-\frac{i}{M\sqrt{\eta^2-Q^2}}\int_C dr(1-r)
\int^\infty_0dx^0(x^0)^2\nonumber\\
&&\times\exp[-x^0/\bar x(\eta,r)]\ 
[Q^2\widetilde{\cal M}_1(x^0,r)+M^2\eta^2\widetilde {\cal M}_2(x^0,r)],
\ear
\bear\label{3.25}
T_2(i\eta,Q^2)&=&\frac{iMQ^2}{\sqrt{\eta^2-Q^2}}\int_C dr(1-r)
\int^\infty_0dx^0(x^0)^2\nonumber\\
&&\times\exp[-x^0/\bar x(\eta,r)]\ \widetilde{\cal M}_2(x^0,r).\ear

We note that for given $Q^2$ and $\eta>Q$ the characteristic
damping length $\bar x(\eta,r)$ is positive for
\be\label{3.26}
{\rm Re}\ r>-\bar r.\ee
We will thus take in the following the curve $C$ in
(\ref{3.24}) and (\ref{3.25}) to be to the right of $-\bar r$ as
shown in Fig. 4. In this way we get 
integral representations of 
$T_{1,2}$
$(i\eta,Q^2)$ where the integration path avoids
 the singularities of the integrand at $r=0,2$, 
which correspond to the light-cone.

The representations (\ref{3.24}) and (\ref{3.25}) will be used as 
basis for the discussion in the following sections.

Let us now estimate the relevant integration range in $r$ in 
(\ref{3.24}) and (\ref{3.25}) for $\eta\to\infty$. We see from (\ref{3.22})
that for fixed $r$ we have
\be\label{3.32}
\bar x(\eta,r)\sim \eta^{-1}\quad {\rm for}\quad \eta\to\infty.\ee
Then the factors $\exp(-x^0/\bar x(\eta,r))$ will suppress such
contributions to the integrals (\ref{3.24}) and (\ref{3.25}).
Now we can keep the curve $C$ in Fig. 4 at finite fixed values 
of $r$ except for the region between $r=0$ and $r=-\bar r$ where $C$
has to cross the negative real axis. Taking as a typical value
$r=-\bar r/2$ we get
\be\label{3.33}
\bar x(\eta,-{\textstyle\frac{1}{2}}\bar r)=\frac{2(1+\bar r)}{\eta\bar r}
\sim\frac{2\eta}{Q^2}\quad {\rm for}\quad \eta\to\infty.\ee
Thus, for $\eta\to\infty$ this region, where $\bar x(\eta,r)$ becomes
very large, will give the main contribution to the integrals
(\ref{3.24}) and (\ref{3.25}). The behaviour of
$T_{1,2}(i\eta,Q^2)$ for $\eta\to\infty$ is therefore expected to 
be governed by the behaviour of $\widetilde{\cal M}_{1,2}(x^0,r)$ for small 
$|r|$ and large $x^0$. 

\section{Effective Hamiltonians and Lagrangians}

\setcounter{equation}{0}
In this section we shall express the matrix elements ${\cal M}_{\mu\nu}$
(\ref{3.5}) as functional integrals with effective Lagrangians containing
$r$ (\ref{3.12a}) as parameter. The procedure is analogous
to the one of \cite{12} but the vector nature of the electromagnetic
current causes some complications. We work in this section with
the Lagrangian of QCD in a general covariant gauge. The case of the
temporal gauge is treated in appendix C.

\subsection{Matrix elements in Minkowski and Euclidean space}

We start here with the following matrix elements in the nucleon
rest frame, supposing always $x^3=x^0(1-r),x^0\geq 0$:
\be\label{4.1}
\widetilde{\cal M}^-_a(x^0,r)={\textstyle {\textstyle\frac{1}{2}}}
\sum_{spins}\langle N(p)
|(-g^{\mu\nu})J_\mu(x^3\vec e_3,x^0)
J_\nu(0)|N(p)\rangle,\ee
\be\label{4.2}
\widetilde{\cal M}^-_b(x^0,r)={\textstyle{\textstyle\frac{1}{2}}}
\sum_{spins}\langle N(p)
|J_0(x^3\vec e_3,x^0)J_0(0)|N(p)\rangle.\ee
The relation of $\widetilde{\cal M}^-_{a,b}(x^0,r)$ and 
$\widetilde {\cal M}^-_{1,2}(x^0,r)$ (\ref{3.12b}) is given in
appendix A. From translational invariance we have
\be\label{4.3}
J_\mu(x^3\vec e_3,x^0)=\exp(ix^0 H_r)
J_\mu(0)\exp(-ix^0 H_r),\ee
where
\be\label{4.4}
H_r=P^0-(1-r)P^3\ee
and $P^0,P^3$ are the energy and third component of the  momentum 
operator. Note that  $H_r$ is positive
semidefinite for $0<r<2$. From (\ref{4.3}) we get
\bear\label{4.5}
\widetilde{\cal M}_a^-(x^0,r)&=&{\textstyle {\textstyle\frac{1}{2}}}
\sum_{spins}\langle N(p)|(-g^{\mu\nu})
\exp(ix^0 H_r)
J_\mu(0)\exp(-ix^0 H_r)J_\nu(0)|N(p)\rangle,\nonumber\\
&&{}\\
\label{4.6}
\widetilde{\cal M}_b^-(x^0,r)&=&{\textstyle {\textstyle\frac{1}{2}}}
\sum_{spins}\langle N(p)|
\exp(ix^0 H_r)
J_0(0)\exp(-ix^0 H_r)J_0(0)|N(p)\rangle .\nonumber\\
&&{}\ear
We see that in the effective theory
$\widetilde{\cal M}^-_{a,b}(x^0,r)$ are correlation functions of
two currents at purely timelike separation $x^0$ .

We will now keep $r$ fixed with $0<r<2$ where $H_r$ is positive
semidefinite. This allows us to continue $\widetilde{\cal
M}^-_{a,b}(x^0,r)$ 
analytically into the lower half $x^0$-plane and in particular
to $x^0=-iX_4$ with $X_4>0$:
\bear\label{4.7}
\widetilde{\cal M}^-_a(-iX_4,r)&=&{\textstyle{\textstyle\frac{1}{2}}}\sum_{spins}
\langle N(p)|(-g^{\mu\nu})\exp\left(X_4 H_r\right)J_\mu(0)
\exp\left(-X_4 H_r\right)J_\nu(0)|N(p)\rangle ,\nonumber\\
&&{}\\
\label{4.8}
\widetilde{\cal M}_b^-(-iX_4,r)&=&{\textstyle{\textstyle\frac{1}{2}}}
\sum_{spins}
\langle N(p)|\exp\left(
X_4 H_r\right)J_0(0)
\exp\left(-X_4 H_r\right)J_0(0)|N(p)\rangle.\nonumber\\
&&{}\ear
The matrix elements (\ref{4.7}), (\ref{4.8}) are correlation functions
of two currents at purely timelike distance $X_4$ in a Euclidean
field theory with $H_r$ as Euclidean Hamiltonian.

In the following we shall write the matrix elements (\ref{4.5})
and (\ref{4.6}) as path integrals in an effective, $r$-dependent, 
Minkoswkian theory and (\ref{4.7}) and (\ref{4.8}) in an
effective Euclidean theory.
\vspace*{.5cm}

\subsection{Effective Lagrangian and path integral in Minkowski space}

We start with the Lagrangian of QCD in a general covariant gauge
\bear\label{4.9}
{\cal L}&=&-{\textstyle\frac{1}{4}} G^a_{\mu\nu}G^{a\mu\nu}
-{\textstyle\frac{1}{2}}\xi^{-1}(\partial^\mu G_{\mu}^a)(\partial^\nu
G_\nu^a)+\sum_q\bar q  
({\textstyle\frac{i}{2}}\gamma^\mu
\stackrel{\leftrightarrow}{D}_\mu-m_q)q\nonumber\\
&&+(\partial^\mu\bar\phi^a)(\partial_\mu \phi^a)
-gf_{abc}(\partial^\mu\bar\phi^a)G^b_\mu\phi^c.\ear
Here $q$ denotes the quark fields, $\phi^a$ the Fadeev-Popov fields,
$G_\mu^a$ the gluon potentials and $G^a_{\mu\nu}$ the gluon field strength
tensor,
\be\label{4.10}
G^a_{\mu\nu}=\partial_\mu G_\nu^a-\partial_\nu G^a_\mu-gf_{abc}G^b_\mu
G^c_\nu.\ee
The coupling constant is $g$, the quark masses are $m_q$, and $D_\mu$
is the covariant derivative
\be\label{4.11}
D_\mu q=(\partial_\mu+ig G^a_\mu{\textstyle\frac{1}{2}}
\lambda_a)q.\ee
All quantities in (\ref{4.9}) are the unrenormalised ones. 

It is now a straightforward exercise to derive from
(\ref{4.9}) the canonical momenta $\Pi$, the Hamiltonian density,
the Hamiltonian $P^0$, the third component of the momentum
$P^3$ and the effective Hamiltonian $H_r$ (\ref{4.4}). The
details are given in appendix B.  Here we list the result:
\be\label{4.12}
H_r=\int d^3x {\cal H}_r,\ee
\bear\label{4.13}
{\cal H}_r&=&-{\textstyle\frac{1}{2}}\xi\Pi_{G^{a0}}\Pi_{G^{a0}}\nonumber\\
&&-\Pi_{G^{a0}}(\partial_j G^{aj}-(1-r)\partial_3 G^{a0})\nonumber\\
&&+{\textstyle\frac{1}{2}} \Pi_{G^{aj}}\Pi_{G^{aj}}\nonumber\\
&&+\Pi_{G^{aj}}(-\partial_j G^{a0}+gf_{abc}G^{b0}G^{cj}+
(1-r)\partial_3 G^{aj})\nonumber\\
&&+{\textstyle\frac{1}{4}}G^a_{jk}G^{ajk}\nonumber\\
&&+\sum_q\bar q(-{\textstyle\frac{i}{2}}\gamma^j
\stackrel{\leftrightarrow}{\partial}_j+
{\textstyle\frac{i}{2}}\gamma^0(1-r)\stackrel{\leftrightarrow}{\partial}_3
+g\gamma^\mu G^a_\mu{\textstyle\frac{1}{2}}\lambda_a+m_q)q\nonumber\\
&&+\Pi_{\phi^a}\Pi_{\bar\phi^a}+\Pi_{\phi^a}(gf_{abc}G^{b0}\phi^c
+(1-r)\partial_3\phi^a)\nonumber\\
&&+(1-r)(\partial_3\bar\phi^a)\Pi_{\bar\phi^a}+(\partial_j\bar\phi^a)\partial_j
\phi^a\nonumber\\
&&+gf_{abc}(\partial_j\bar\phi^a) G^{bj}\phi^c.\ear
Here and in the following Latin indices $j,k,...$ run from 1 to 3. From the
effective Hamiltonian density (\ref{4.13}) we get in the 
standard way the effective Lagrangian density:
\bear\label{4.14}
{\cal L}_r&=&-{\textstyle\frac{1}{4}}G^a_{\mu\nu}G^{a\mu\nu}
-(1-r)G^{a0j}\partial_3 G^{aj}
+{\textstyle\frac{1}{2}}(1-r)^2(\partial_3 G^{aj})\partial_3 G^{aj}\nonumber\\
&&-{\textstyle\frac{1}{2}}\xi^{-1}
(\partial_\mu G^{a\mu}-(1-r)\partial_3 G^{a0})
(\partial_\nu G^{a\nu}-(1-r)\partial_3 G^{a0})\nonumber\\
&&+\sum_q\bar q({\textstyle\frac{i}{2}}\gamma^\mu
\stackrel{\leftrightarrow}{D}_\mu-m_q-(1-r){\textstyle\frac{i}{2}}
\gamma^0\stackrel{\leftrightarrow}{\partial_3})q\nonumber\\
&&+(\partial^\mu\bar\phi^a)\partial_\mu\phi^a-gf_{abc}
(\partial_\mu\bar\phi^a)G^{b\mu}\phi^c\nonumber\\
&&-(1-r)\left((\partial_3\bar\phi^a)\dot\phi^a+
\dot{\bar \phi^a}\partial_3\phi^a\right)\nonumber\\
&&+(1-r)\left(\partial_3\bar\phi^a\right) gf_{abc}G^{b0}\phi^c\nonumber\\
&&+(1-r)^2\left(\partial_3\bar\phi^a\right) \partial_3\phi^a.\ear

Here and in the following we use the term ``effective'' for the
$r$-dependent theory. We emphasize that our procedure does not imply
approximations like integrating out some modes etc.  The $r$-dependent
theory is as good as the original one for the calculation of our
matrix elements and we hope that it will be more effective for the
study of the small $x_{Bj}$ limit of the structure functions.

We can now give the path integral representation of $
\widetilde{\cal M}^-_{a,b}(x^0,r)$ (\ref{4.5}), (\ref{4.6}) in the 
effective theory described by (\ref{4.14}).
The procedure is analogous to the one described in detail in
\cite{12}.
 Let $\psi_N(x)$ be an interpolating field operator for the nucleon
with normalisation such that
\be\label{4.15}
\langle 0|\psi_N(x)|N(p,s)\rangle =e^{-ipx}u_s(p),\qquad
(s=\pm1/2).\ee
According to the LSZ formalism \cite{19} we define
\be\label{4.16}
A_s(p,t)=\int_{x^0=t} d^3x\ \bar u_s(p)e^{ipx}\gamma^0\psi_N(x).
\ee
In the sense of the weak limit we have
\bear\label{4.17}
\lim_{t\to-\infty}A_s(p,t)&=&A^{\rm in}_s(p),\nonumber\\
\lim_{t\to+\infty}A_s(p,t)&=&A_s^{\rm out} (p),
\ear
where $A_s^{\rm in}(p)$ and $A_s^{\rm out}(p)$ are the annihilation operators for
incoming and outgoing nucleons. In the standard way we obtain now
for the matrix elements
\bear\label{4.18}
\widetilde {\cal M}_a^-(x^0,r)&=&\lim_{t_i\to-\infty \atop t_f\to +\infty}
\exp[iM(t_f-t_i)]\nonumber\\
&&\times Z^{-1}\int{\cal D}(G,q,\bar q, \phi,\bar\phi
)\exp[i\int d^4 x{\cal L}_r(x)]\nonumber\\
&&\times {\textstyle\frac{1}{2}}\sum_s a_s(p,t_f)
(-g^{\mu\nu})j_\mu(\vec 0,x^0)j_\nu
(0)a^\dagger_s(p,t_i),\\
\label{4.19}
Z&=&\int{\cal D}(G,q,\bar q,\phi,\bar\phi)\exp[i\int d^4x{\cal L}_r(x)],\\
\label{4.20}
\widetilde{\cal M}_b^-(x^0,r)&=&\lim_{t_i\to-\infty \atop t_f\to +\infty}
\exp[iM(t_f-t_i)]\nonumber\\
&&\times Z^{-1}\int {\cal D}(G,q,\bar q,\phi,\bar\phi
)\exp[i\int d^4x{\cal L}_r(x)]\nonumber\\
&&\times {\textstyle\frac{1}{2}}\sum_s a_s(p,t_f)j_0
(\vec 0,x^0)j_0(0)a^\dagger_s
(p,t_i).\ear
Here $p$ is always the nucleon momentum in the rest system (see (\ref{2.14}))
and $a_s,a_s^\dagger, j_\mu$ are obtained by replacing in $A_s,A^\dagger_s,J_\mu$ the quark field operators by the corresponding
Grassmann variables. Of course, also in ${\cal L}_r(x)$ quark,
ghost and gluon field operators have to be replaced by Grassmann
variables and classical gluon fields, respectively.
With (\ref{4.18})-(\ref{4.20}) we have the  path integral
representation of the matrix elements $\widetilde{\cal M}^-_{a,b}$ as
timelike correlation functions of the currents in the effective,
$r$-dependent Minkowskian theory.

\subsection{Effective Lagrangian and path integral in Euclidean space}

Here we derive the path integral representation for the matrix elements
(\ref{4.7}), (\ref{4.8}) in the Euclidean effective theory. Points in
Euclidean space are denoted by $X=(\vec X,X_4)$. Our Euclidean
$\gamma$-matrices are
\bear\label{4.21}
&&\gamma_{Ej}=-i\gamma^j,\quad (j=1,2,3),\nonumber\\
&&\gamma_{E4}=\gamma^0.\ear
We perform now a  rotation to Euclidean space starting from the
effective Hamiltonian (\ref{4.12}), (\ref{4.13}). The details are
given in appendix D. The result for the effective Lagrangian density
in Euclidean space is:
\bear\label{4.22}
{\cal L}_{E,r}&=&{\textstyle\frac{1}{2}}\xi^{-1}[\partial_\mu G^a_{E\mu}
+i(1-r)\partial_3G^a_{E4}][\partial_\nu G^a_{E\nu}
+i(1-r)\partial_3G^a_{E4}]\nonumber\\
&&+{\textstyle\frac{1}{2}}[G^a_{E4j}+i(1-r)
\partial_3G^a_{Ej}]
[G^a_{E4j}+i(1-r)\partial_3G^a_{Ej}]\nonumber\\
&&+{\textstyle\frac{1}{4}}G^a_{Ejk}G^a_{Ejk}\nonumber\\
&&+\sum_q\bar q_E\left({\textstyle\frac{1}{2}}\gamma_{ E\mu}
\stackrel{\leftrightarrow}{\partial}_\mu+{\textstyle\frac{i}{2}}
\gamma_{E4}(1-r)
\stackrel{\leftrightarrow}{\partial}
_3+ig\gamma_{E\mu}G^a_{E\mu}{\textstyle\frac{1}{2}}
\lambda_a+m_q\right)q_E\nonumber\\
&&+(\partial_4\bar\phi^a_E+i(1-r)\partial_3\bar\phi^a_E)(\partial
_4\phi^a_E-gf_{abc}G^b_{E4}\phi^c_E+i(1-r)\partial_3\phi^a_E),\nonumber\\
&&+(\partial_j\bar\phi_E^a)\partial_j\phi^a_E-gf_{abc}
(\partial_j\bar\phi^a_E)G^b_{Ej}\phi^c_E.\ear

For the following it is convenient to split ${\cal L}_{E,r}$ into
the quadratic and the interaction term:
\be\label{4.23}
{\cal L}_{E,r}={\cal L}^{(0)}_{E,r}+{\cal L}^{\rm Int}
_{E,r},\ee
\bear\label{4.24}
{\cal L}_{E,r}^{(0)}&=&{\textstyle\frac{1}{2}}\xi^{-1}
[\partial_\mu G^a_{E\mu}
+i(1-r)\partial_3G^a_{E4}]
[\partial_\nu G^a_{E\nu}
+i(1-r)\partial_3G^a_{E4}]\nonumber\\
&&+{\textstyle\frac{1}{2}}[\partial_4G^a_{Ej}-
\partial_jG^a_{E4}+i(1-r)\partial_3G^a_{Ej}]
[\partial_4G^a_{Ej}-\partial_jG^a_{E4}
+i(1-r)\partial_3G^a_{Ej}]\nonumber\\
&&+{\textstyle\frac{1}{4}}[\partial_jG^a_{Ek}
-\partial_kG^a_{Ej}][\partial_jG^a_{Ek}-\partial_kG^a_{Ej}]
\nonumber\\
&&+\sum_q\bar q_E[{\textstyle\frac{1}{2}}\gamma_{E\mu}
\stackrel{\leftrightarrow}{\partial}
_\mu+{\textstyle\frac{i}{2}}\gamma_{E4}(1-r)
\stackrel{\leftrightarrow}{\partial}_3+m_q]q_E\nonumber\\
&&+(\partial_4\bar\phi^a_E+i(1-r)\partial_3\bar\phi^a_E)(\partial
_4\phi^a_E+i(1-r)\partial_3\phi^a_E)\nonumber\\
&&+(\partial_j\bar\phi_E^a)(\partial_j\phi^a_E),
\ear
\bear\label{4.25}
{\cal L}_{E,r}^{{\rm Int}}&=&-gf_{abc}[\partial_4G^a_{Ej}-\partial_j
G^a_{E4}+i(1-r)
\partial_3G^a_{Ej}]G^b_{E4}G^c_{Ej}\nonumber\\
&&-{\textstyle\frac{1}{2}}gf_{abc}(\partial_jG^a_{Ek}-
\partial_kG^a_{Ej})G^b_{Ej}G^c_{Ek}\nonumber\\
&&+{\textstyle\frac{1}{4}}g^2f_{abc}f_{ab'c'}G^b_{E\mu}
G^c_{E\nu}G^{b'}_{E\mu}G^{c'}_{E\nu}\nonumber\\
&&+ ig\sum_q\bar q_E\gamma_{E\mu}G^a_{E\mu}{\textstyle\frac{1}{2}}
\lambda_aq_E\nonumber\\
&&-gf_{abc}[(\partial_\mu\bar\phi^a_E)G^b_{E\mu}+i(1-r)
(\partial_3\bar\phi^a_E)G^b_{E4}]\phi^c_E\,.\ear
With the same procedure as in \cite{12} we get the following path
integral representation for the matrix elements $\widetilde{\cal M}
^-_{a,b}(-iX_4,r)$ of (\ref{4.7}) and (\ref{4.8}):
\bear\label{4.26}
\widetilde{\cal M}_a^-(-iX_4,r)&=&{\textstyle
\frac{1}{2}}\sum_s\lim_{\tau_i\to-\infty,
\atop\tau_f\to+\infty}
\exp[(\tau_f-\tau_i)M]\ Z^{-1}_E
\int {\cal D}(G_E,q_E,\bar q_E,\phi_E,\bar\phi_E)\nonumber\\
&&\times  a_{s,E}(p,\tau_f)
(-1)j_{E\mu}(\vec 0,X_4)j_{E\mu}(0)a^\dagger_s
(p,\tau_i)\exp(-S_{E,r}),\ear
\bear\label{4.27}
\widetilde{\cal M}_b^-(-iX_4,r)&=&{\textstyle
\frac{1}{2}}\sum_s\lim_{\tau_i\to-\infty,
\atop\tau_f\to\infty}
\exp[(\tau_f-\tau_i)M]\ Z^{-1}_E
\int {\cal D}(G_E,q_E,\bar q_E,\phi_E,\bar\phi_E)\nonumber\\
&&\times a_{s,E}(p,\tau_f)
j_{E,4}(\vec 0,X_4)j_{E4}(0)a^\dagger_s(p,\tau_i)
\exp(-S_{E,r}).\ear
Here
\be\label{4.28}
Z_E=\int D(G_E,q_E,\bar q_E,\phi_E,\bar\phi_E)\ \exp(-S_{E,r}),\ee
\be\label{4.29}
S_{E,r}=\int d^4X\ {\cal L}_{E,r}(X),\ee
$G_E,...,\bar\phi_E$ are the integration variables,
$j_{E\mu}$ represent the components of the electromagnetic current and
$a_s,a_s^\dagger$ the nucleon state; see appendix D.

From (\ref{4.23})-(\ref{4.25}) we see that the Euclidean, $r$-dependent action $S_{E,r}$ has, in general, imaginary parts. To discuss the formal
convergence properties of the integrals over the gluon-potential
variables we split $S_{E,r}$ into the quadratic and interaction parts
\be\label{4.30}
S_{E,r}=S_{E,r}^{(0)}+S_{E,r}^{\rm Int},\ee
\be\label{4.31}
S_{E,r}^{(0)}=\int d^4X{\cal L}^{(0)}_{E,r},\ee
\be\label{4.32}
S_{E,r}^{\rm Int}=\int d^4X{\cal L}^{\rm Int}_{E,r}.\ee
After some partial integrations we get
\bear\label{4.33}
S^{(0)}_{E,r}&=&\int d^4X\Big\{{\textstyle\frac{1}{2}} G^a_{E\mu}\delta_{ab}
\Big[-\delta_{\mu\nu}\partial_\lambda\partial_\lambda+(1-
\xi^{-1})\partial_\mu\partial_\nu-2i(1-r)\delta_{\mu\nu}
\partial_3\partial_4\nonumber\\
&&+i(1-r)(1-\xi^{-1})(\delta_{\mu4}\partial_\nu\partial_3+
\delta_{\nu4}\partial_\mu\partial_3)\nonumber\\
&&+(1-r)^2\delta_{\mu
\nu}\partial_3\partial_3-(1-r)^2
(1-\xi^{-1})\delta_{\mu4}\delta_{\nu4}\partial_3\partial_3\Big]
G^b_{\nu E}\nonumber\\
&&+\sum_q\bar q_E[\gamma_{E\mu}\partial_\mu+i(1-r)\gamma_{E4}\partial_3+m_q]q_E
\nonumber
\\
&&+\bar\phi^a_E\delta_{ab}(-\partial_\lambda\partial_\lambda-2i(1-r)
\partial_3\partial_4+(1-r)^2\partial_3\partial_3)\phi^b_E\Big\}.
\ear
It is particularly convenient to use the Feynman-'t Hooft gauge,
that is to set $\xi=1$ in the following. With this we find
\bear\label{4.34}
S^{(0)}_{E,r}|_{\xi=1}&=&\int d^4X\Big\{{\textstyle\frac{1}{2}}G^a_{E\mu}\delta_{ab}\delta_{\mu\nu}
[-\partial_\lambda\partial_\lambda-2i(1-r)
\partial_3\partial_4+(1-r)^2\partial_3\partial_3]G^b_{E\nu}
\nonumber\\
&&+\sum_q\bar q_E[\gamma_{E\mu}\partial_\mu+i(1-r)\gamma_{E4}\partial_3
+m_q]q_E\nonumber\\
&&+\bar\phi^a_E\delta_{ab}[-\partial_\lambda\partial_\lambda
-2i(1-r)\partial_3\partial_4+(1-r)^2\partial_3\partial_3]\phi^b_E\Big\}.
\ear
With the Fourier transformations
\bear\label{4.35}
G^a_{E\mu}(X)&=&\int{\textstyle\frac{d^4K}{(2\pi)^4}}e^{iKX}\widetilde G^a
_{E\mu}(K),\nonumber\\
(\widetilde G^a_{E\mu}(K))^*&=&\widetilde G^a_{E\mu}(-K),\nonumber\\
q_E(X)&=&\int{\textstyle\frac{d^4K}{(2\pi)^4}}e^{iKX}\widetilde q_E
(K),\nonumber\\
\phi^a_E(X)&=&\int{\textstyle\frac{d^4K}{(2\pi)^4}}e^{iKX}\widetilde \phi_E^a
(K),\nonumber\\
\overline{\phi^a_E}(X)&=&\int{\textstyle\frac{d^4K}{(2\pi)^4}}e^{-iKX}
\widetilde{\bar \phi^a}
(K)\ear
we get
\bear\label{4.36}
S^{(0)}_{E,r}|_{\xi=1}&=&\int{\textstyle\frac{d^4K}{(2\pi)^4}}\Big\{
{\textstyle\frac{1}{2}}(\widetilde G^a_{E\mu}(K))^*\delta_{ab}\delta_{\mu\nu}
\nonumber\\
&&[K^2+2i(1-r)K_3K_4-(1-r)^2K^2_3]\widetilde G^b_{E\nu}(K)\nonumber\\
&&+\sum_q\bar{\tilde q}_E(K)[i\gamma_{E\mu}K_\mu-(1-r)\gamma_{E4}K_3+m_q]
\tilde q_E(K)\nonumber\\
&&+\widetilde{\bar\phi^a}(K)\delta_{ab}[K^2+2i(1-r)K_3K_4-(1-r)^2K^2_3]
\tilde\phi^b(K)\Big\}.\ear
Note that
\bear\label{4.37}
&&K^2+2i(1-r)K_3K_4-(1-r)^2K^2_3\nonumber\\
&=&K^2_1+K^2_2+r(2-r)K^2_3+K^2_4+2i(1-r)K_3K_4\ear
has a positive real part for $0<r<2$ and $K\not=0$.

We discuss now the formal convergence properties of the gluonic
integrations in (\ref{4.26})-(\ref{4.28}), setting $\xi=1$.
Consider a given gluon potential $G_{E\mu}^a(X)$ together with 
$\lambda G^a_{E\mu}(X)$ for all $\lambda>0$, to investigate
how the integrand in the functional integral behaves for large 
potentials. Suppose first that 
\be\label{4.38}
f_{abc}G^b_{E\mu}(X)G^c_{E\nu}(X)\not\equiv0.\ee
Then the term in ${\cal L}^{\rm Int}_{E,r}$ (\ref{4.25}) quartic in
the gluon potentials ensures 
\be\label{4.39}
\exp[-S_{E,r}(\lambda G)]\propto e^{-\lambda^4 c_1}
\quad {\rm for}\quad \lambda\to\infty\ee
with $c_1>0$. On the other hand, if 
\be\label{4.39a}
f_{abc}G^b_{E\mu}(X)G^c_{E\nu}(X)=0\ee
for all $X$, the positivity of the real part of the quadratic term of 
the action (\ref{4.36}) ensures that
\be\label{4.40}
\exp[-S_{E,r}(\lambda G)|_{\xi=1}]\propto e^{-\lambda^2 c_2}
\quad {\rm for}\quad \lambda\to\infty,\ee
with $c_2>0$. Thus in any case the integrand of the functional integrals
(\ref{4.26})-(\ref{4.28}) is damped exponentially for large gluon
potentials. This should make the integrals well behaved after 
introducing some regularisation procedure, for instance
a lattice regularisation.

\subsection{Propagators in Euclidean space and perturbation expansion}

In this section we discuss first the lowest-order propagators in
Euclidean space for the gauge choice $\xi=1$. The basic Green's
function  $\Delta_E(X,m^2,r)$
for mass $m$ in the $r$-dependent theory is defined through
\be\label{4.42}
[-\partial_\lambda\partial_\lambda-2i(1-r)\partial_3\partial_4+
(1-r)^2\partial_3\partial_3+m^2]
\Delta_E(X,m^2,r)=\delta^4(X).\ee
The solution of (\ref{4.42}) as given in \cite{12} is
\bear\label{4.43}
\Delta_E(X,m^2,r)&=&\int\frac{d^4K}{(2\pi)^4}e^{iKX}(K^TA_rK+m^2)^{-1}
\nonumber\\
&=&\frac{m}{4\pi^2}(X^TA_r^{-1}X)^{-1/2}K_1
\bigl(m(X^TA^{-1}_rX)^{1/2}\bigr),
\ear
where $K_1$ is the modified Bessel function of order 1 and
\be\label{4.44}
A_r=\left(\begin{array}{cccc}
1&0&0&0\\
0&1&0&0\\
0&0&r(2-r)&i(1-r)\\
0&0&i(1-r)&1\end{array}\right),\ee
\be\label{4.45}
A^{-1}_r=\left(\begin{array}{cccc}
1&0&0&0\\
0&1&0&0\\0&0&1&-i(1-r)\\
0&0&-i(1-r)&r(2-r)\end{array}\right).\ee
For $m^2=0$ we get
\bear\label{4.46}
\Delta_E(X,0,r)&=&\frac{1}{4\pi^2}(X^TA^{-1}_rX)^{-1}\nonumber\\
&=&\frac{1}{4\pi^2}
\bigl(X^2_1+X^2_2+X^2_3+r(2-r)X^2_4-2i(1-r)X_3X_4\bigr)^{-1}.\nonumber\\
&&{}\ear

From (\ref{4.34}) we get now easily the expressions for the propagators
in lowest order. The gluon propagator is
\be\label{4.47}
\Delta_{E\mu\nu}^{(0)ab}(X,r)=\delta_{ab}\delta_{\mu\nu}\Delta_E(X,0,r),\ee
the quark propagator is
\be\label{4.48}
S_E^{(0)}(X,m_q,r)=[-\gamma_{E\mu}\partial_\mu-i(1-r)\gamma_{E4}
\partial_3+m_q]\Delta_E(X,m^2_q,r),\ee
and the ghost propagator is
\be\label{4.49}
\Delta_E^{(0)ab}(X,r)=\delta_{ab}\Delta_E(X,0,r).\ee
All these propagators (\ref{4.43}), (\ref{4.47})-(\ref{4.49}) have the
property that for $r\to 0$ they fall off more and more slowly in
$X_4$ direction, whereas their fall-off in $X_1,X_2$ and $X_3$
directions is independent of $r$. For the case $m\not=0$ this is
discussed in section 5 of \cite{12}. For the propagator of a quark
of mass $m_q$ the correlation length in the directions $X_j(j=1,2,3)$ is
\be\label{4.50}
l^{(0)}_T=m^{-1}_q,\ee
and in the direction $X_4$
\bear\label{4.51}
l_4^{(0)}&=&[m_q^2r(2-r)]^{-1/2}\nonumber\\ 
&\approx& l_T^{(0)}(2r)^{-1/2}\quad {\rm for}\ r\to 0.\ear
For the massless case (\ref{4.46}) there is of course no 
genuine correlation length, but for $r\to 0$ the fall-off
in $X_4$ direction is again slower by a factor $(2r)^{-1/2}$ compared
to the $X_{1,2,3}$  directions.

We can now use the propagators (\ref{4.47})-(\ref{4.49}) to construct
the unrenormalised perturbation series for the Green's functions 
in the theory described by ${\cal L}_{E,r}$ (\ref{4.22}). Since the Fourier
transforms of the propagators (\ref{4.43}), (\ref{4.47})-(\ref{4.49})
have no unwanted poles in momentum space, the convergence
properties of Feynman integrals should be the same for arbitrary $r$ as for
the standard case $r=1$. Thus we conclude that also the construction of the 
renormalised perturbation series should work for the ${\cal L}_{E,r}$ theory
as for the standard case. Of course, the purpose of our paper is not
to advocate perturbative calculations starting with the Lagrangian
density ${\cal L}_{E,r}$ (\ref{4.23}) but to
provide a framework for nonperturbative calculations. 

\section{Phenomenological applications}

In this section we study the naive parton model from our point
of view and make some speculations concerning a possible critical
behaviour of the full theory with interaction for $r\to 0$.

\subsection{The naive parton model}
\setcounter{equation}{0}

The central assumption of the naive parton model is that of free
field or canonical behaviour of the product of currents near the
light cone. In detail one assumes \cite{18} for ${\cal M}_{1,2}$
of (\ref{3.5})
\bear\label{5.1}
{\cal M}_1(xp,x^2)&\sim& h_1(xp)\frac{4\pi^2}{i}\Delta^+(x,m^2)\nonumber\\
&\sim& h_1(xp)(-x^2+i\epsilon(xp))^{-1},\\
\label{5.2}
{\cal M}_2(xp,x^2)&\sim&-h_2(xp)
8\pi^2\int^\infty_0ds\delta'(s-m^2){\textstyle\frac{1}{i}}
\Delta^+(x,s)\nonumber\\
&\sim&{\textstyle\frac{1}{2}}h_2(xp)\ln[m^2(-x^2+i\epsilon(xp))],
\ear
where $\sim$ indicates that only the leading term for $x^2\to 0$ is
considered and $\Delta^+$ is defined in (\ref{3.14}).
The parameter $m$ represents a hadronic mass scale. Its precise value 
does not matter except that we will assume
\be\label{5.3}
0<m\leq M.\ee
According to the DGS representations (\ref{3.18b}), 
(\ref{3.18c}) the functions $h_{1,2}(xp)$ can be represented as
\be\label{5.4}
h_1(xp)=\int^1_{-1}d\zeta[\cos(\zeta px)-1]\tilde h_1(\zeta)
+h_1^{(0)},\ee
\be\label{5.5}
h_2(xp)=\int^1_{-1}d\zeta\cos(\zeta px)\tilde h_2(\zeta),\ee
where
\bear\label{5.6}
\tilde h_j(\zeta)&=&\tilde h_j(-\zeta)=\tilde h_j^*(\zeta),
\qquad(j=1,2),\nonumber\\
h_1^{(0)}&=&h_1^{(0)*}={\rm const.}\ear

It is now an easy exercise to calculate the structure functions 
$W_{1,2}$ inserting the parton model ansatz (\ref{5.1})-(\ref{5.6})
into (\ref{3.8}), (\ref{3.9}). The result, expressed in terms of
$F_{2,L}$ of (\ref{2.7a}), is
\bear\label{5.7}
F_2(x_{Bj},Q^2)&=&\nu W_2(\nu,Q^2)
\sim 2\pi^2 x_{Bj}\tilde h_2'(x_{Bj}),\\
\label{5.8}
F_L(x_{Bj},Q^2)&\sim& 4\pi^2x^2_{Bj}\tilde h_1(x_{Bj}),\ear
where now $\sim$ indicates the leading term for $Q^2\to\infty$.
Of course, we get Bjorken scaling. From the well-known relations
of the parton model we get the physical interpretation of the
functions $\tilde h_1$ and $\tilde h_2$ as
\bear\label{5.9}
2\pi^2\tilde h_2'(\zeta)&=&\sum_je^2_jN_j(\zeta)+
\sum_j\tilde e_j^2\tilde N_j(\zeta),\nonumber\\
4\pi^2\tilde h_1(\zeta)&=&\zeta^{-1}\sum_j\tilde e_j^2\tilde N_j(\zeta),
\nonumber\\
\qquad &&0<\zeta\leq 1,\ear
where $e_j,N_j(\zeta)$ are the charges and distribution functions
for the spin 1/2 partons and $\tilde e_j,\tilde N_j(\zeta)$ for the spin
0 partons. 

Using an approximate parton model description, we conclude from
experiment \cite{1}-\cite{3} and the fits in \cite{317} that the growth
of $F_2$ for $x_{Bj}\to 0$ is at most as
\be\label{5.10}
F_2(x_{Bj},Q^2)\propto (x_{Bj})^{-a}\ee
with $0<a<1$. This implies from (\ref{5.7}) that at most
\bear\label{5.11}
\tilde h_2'(\zeta)&\propto &|\zeta|^{-a-1},\nonumber\\
\tilde h_2(\zeta)&\propto &|\zeta|^{-a},\ear
for $\zeta\to 0$. The structure function $F_L$ is not well measured
at high $Q^2$. All indications are that $F_L$ is
small compared to $F_2$. Thus
a conservative estimate is that also $F_L$ grows at most as
\be\label{5.12}
F_L(x_{Bj},Q^2)\propto (x_{Bj})^{-a}\ee
for $x_{Bj}\to 0$. From (\ref{5.8}) this implies that at most
\be\label{5.13}
\tilde h_1(\zeta)\propto |\zeta|^{-a-2}\ee
for $\zeta\to 0$. We see that with (\ref{5.11}) and (\ref{5.13}) the
integrals (\ref{5.5}) and (\ref{5.4}) are perfectly convergent for $\zeta=0$.
On the other hand, an unsubtracted integral as for $h_2$ would not be 
convergent for $h_1$. These findings are our motivation to 
write the DGS representations (\ref{3.18b}) without, and (\ref{3.18c})
with subtractions.

The purpose of our discussion of the parton model is to see
how one obtains in this case a relation between the behaviour of the structure
functions at large $\nu$, that is small $x_{Bj}$, of the amplitudes
$T_{1,2}$ (\ref{3.24}), (\ref{3.25}) at large $\eta$, that is large
imaginary $\nu$, and of the matrix elements 
$\widetilde {\cal M}_{1,2}$ 
(\ref{3.18}) and $\widetilde{\cal M}_{a,b}^-$ (\ref{4.1}),
(\ref{4.2}) at small $r$. 
For illustration we consider only $W_2,T_2$ and $\widetilde {\cal M}_2$ 
and make the following simple ansatz
\be\label{5.14}
2\pi^2\tilde h_2(\zeta)=-A{\textstyle\frac{1}{a}}(|\zeta|^{-a}-1)(1-
|\zeta|)^b,\ee
with
\be\label{5.15}
0<a<1\ , \quad b>1\ ,\quad A>0.\ee
We get then for $0<\zeta\leq 1$
\bear\label{5.16}
2\pi^2\tilde h_2'(\zeta)&=&
A[1-\zeta+{\textstyle\frac{b}{a}}\zeta(1-\zeta^a)]\zeta^{-a-1}(1-\zeta)^{b-1}
\nonumber\\
&\sim& A\zeta^{-a-1}\qquad {\rm for\ small\ } \zeta\ear
and from (\ref{5.7}) for small $x_{Bj}$
\be\label{5.17}
F_2(x_{Bj},Q^2)\sim A(x_{Bj})^{-a},\ee
which implies for large $\nu$ at fixed $Q^2$:
\be\label{5.18}
W_2(\nu,Q^2)\sim \frac{A}{\nu}\left(\frac{2M\nu}{Q^2}\right)^a,\ee
\be\label{5.18a}
\sigma_T(\nu,Q^2)+\sigma_L(\nu,Q^2)\sim 4\pi^2\alpha
\frac{A}{Q^2}\left(\frac{2M\nu}{Q^2}\right)^a.\ee

On the other hand, we have argued at the end of section 3 that
the behaviour of $T_2(i\eta,Q^2)$ for $\eta\to\infty$ should be
governed by the behaviour of $\widetilde {\cal M}_2(x^0,r)$ (\ref{3.18})
for small $|r|$ and large $x^0$. With the ansatz (\ref{5.2}) we get
\be\label{5.19}
\widetilde{\cal M}_2(x^0,r)\sim{\textstyle\frac{1}{2}}h_2(x^0M)\ln(-r)\ee
for small $|r|$, where from (\ref{5.5}) and (\ref{5.14}) :
\be\label{5.20}
h_2(x^0M)=-\frac{A}{\pi^2a}\int^1_0d\zeta\cos(\zeta x^0M)(\zeta^{-a}-1)
(1-\zeta)^b.\ee
To see the large $x^0$ behaviour of $h_2(x^0M)$, we write it
as follows:
\bear\label{5.21}
h_2(x^0M)&=&-\frac{A}{\pi^2a}\int^\infty_0d\zeta\cos(\zeta x^0M)
\zeta^{-a}e^{-\zeta}\nonumber\\
&&-\frac{A}{\pi^2a}\int^\infty_0d\zeta\cos(\zeta x^0M)
[(\zeta^{-a}-1)(1-\zeta)^b\theta(1-\zeta)-\zeta^{-a}e^{-\zeta}].\nonumber\\
&&{}\ear
The second integral on the r.h.s. of (\ref{5.21}) vanishes faster
than $1/x^0$ for $x^0\to\infty$, as we find from a simple application
of the Riemann-Lebesgue lemma. The first integral  then gives  the
leading behaviour for large $x^0$:
\be\label{5.22}
h_2(x^0M)\sim  -\frac{A}{\pi^2a}\Gamma(1-a)\sin(\frac{\pi}{2}a)(x^0
M)^{a-1}.\ee
Inserting (\ref{5.19}), (\ref{5.22}) into (\ref{3.25}) gives for large
$\eta$:
\bear\label{5.23}
T_2(i\eta,Q^2)&\sim& -\frac{iMQ^2}{2\eta}\frac{AM^{a-1}}{\pi^2a}
\Gamma(1-a)\sin(\frac{\pi}{2}a)\nonumber\\
&&\times \int_C dr\ \ln(-r)\int^\infty_0dx^0(x^0)^{1+a}\exp
(-x^0/\bar x(\eta,r))\nonumber\\
&\sim&-\frac{iMQ^2}{2\eta^{3+a}}\frac{A}{\pi^2a}M^{a-1}
\Gamma(2+a)\Gamma(1-a)\sin({\textstyle\frac{\pi}{2}}a)\nonumber\\
&&\times \int_C dr\ \ln (-r)(r+\bar r)^{-2-a}\nonumber\\
&\sim& A(\cos({\textstyle\frac{\pi}{2}}a))^{-1} \frac
{1}{\eta}\left(\frac{Q^2}{2\eta M}\right)^{-a}.\ear
With analytic continuation we find for large $|\nu|$
\be\label{5.24}
T_2(\nu,Q^2)\sim A(\cos({\textstyle\frac{\pi}{2}}a))^{-1}
[\sin({\textstyle\frac{\pi}{2}}a)+i\cos({\textstyle\frac{\pi}{2}}a)]
\frac{1}{\nu}\left(\frac{Q^2}{2M\nu}\right)^{-a}\ee
and for large real $\nu$
\be\label{5.25}
W_2(\nu,Q^2)={\rm Im}\ T_2(\nu+i\epsilon,Q^2)\sim A\frac{1}{\nu}
\left(\frac{2M\nu}{Q^2}\right)^a.\ee
Of course, this is in perfect agreement with (\ref{5.18}).

We have thus seen in this simple example that the power behaviour
(\ref{5.17}) of $F_2$ corresponds to a behaviour
\be\label{5.26}
\widetilde{\cal M}_2(x^0,r)\propto(x^0M)^{a-1}\ln(-r)\ee
for small $|r|$ and large $x^0$ (see (\ref{5.19}),
(\ref{5.22})).  If we assume that $\widetilde{\cal M}_1(x^0,r)$ has a similar
behaviour, as is for instance true for vanishing $\sigma_L$ (see
(\ref{3.9b})), we get from (\ref{5.26}) and (\ref{A.1}), (\ref{A.2})
for the current correlation functions (\ref{4.7}), (\ref{4.8}):
\be\label{5.27}
\widetilde{\cal M}_{a,b}^-(-iX_4,r)\propto(-iX_4 M)^{a-3}(-r)^{-2}\ee
for small $|r|$ and large $X_4$.

\subsection{Regge behaviour}

We have seen in section 4.4 that the free propagators in the
$r$-dependent Euclidean theory develop in $X_4$ direction
 a large correlation length $\propto 1/\sqrt{r}  $ for $r\to 0$. Let us
assume here that this property remains true in the full theory with
interaction.  In \cite{12} it has been argued that one could then
expect to see a critical behaviour of the theory for $r\to 0$ with $r$
playing the role of the deviation of the temperature from the critical
one, in the statistical physics of a system near a second order phase
transition at $T_c$,
\be\label{5.27a} r\sim (T-T_c)/T_c. \ee
We can then expect to see simple power behaviour of the correlation
functions in $X_4$ direction for $r\to 0$ from general scaling and
renormalization group arguments. In \cite{12} various possibilities
for the behaviour of the correlation length in $X_4$ direction are
discussed.

Let us assume here, as an example, that the $r$-dependent theory of
section 4.3 has a correlation length in $X_4$ direction similar to 
(\ref{4.51})
\be\label{5.28}
l_4(r)\sim m^{-1}(-r)^{-\frac{1}{2}}\ee
for $r\to 0$, where $m$ is a hadronic mass scale. We assume,
furthermore, simple power behaviour of the correlation functions
(\ref{4.7}), (\ref{4.8})
\bear\label{5.29}
&&\widetilde{\cal M}^-_{a,b}(-iX_4,r)
\propto (-iX_4 M)^{a-3}(-r)^{-2-\frac{1}{2}\varepsilon_0},\nonumber\\
&&0\leq a<1,\quad |\varepsilon_0|<1,\ear
for $r\to 0$ and $m^{-1}\ll X_4\ll |l_4(r)|$.
The ansatz (\ref{5.29}) corresponds to (see (A.1), (A.2)):
\bear\label{5.30}
\widetilde{\cal M}_2(x^0,r)&\sim& A'(x^0
M)^{a-1}(-r)^{-\frac{1}{2}\varepsilon_0} ,\nonumber\\
&& A'= {\rm const.}\ear
for $r\to 0$ and $m^{-1}\ll x^0\ll|l_4(r)|$.
What are the consequences of (\ref{5.30}) for $T_2(i\eta,Q^2)$ of
(\ref{3.25}) and the structure function $F_2(x_{Bj},Q^2)$ of
(\ref{2.7a})? We have argued at the end of section 3.3 that the
relevant region of the $r$ integration in (\ref{3.25}) is for
$r\approx -\bar r/2$. Consider now the $x^0$ integral in (\ref{3.25})
for this  value of $r$. We have on the one hand the exponential
damping factor $\exp (-x^0/\bar x(\eta,r))$, on the other hand, the
matrix element $\widetilde{\cal M}_2(x^0,r)$ provides as cutoff the
correlation length $l_4(r)$. For $r=-\bar r/2$ we get from
(\ref{3.23}) and (\ref{5.28}) for  $\eta\gg Q$:
\bear\label{5.31}
\bar r&\approx &\frac{Q^2}{2\eta^2},\nonumber\\
\bar x(\eta,-{\textstyle\frac{1}{2}}\bar r)&\approx
&\frac{4\eta}{Q^2},
\nonumber\\
l_4(-{\textstyle\frac{1}{2}}\bar r)&\approx &\frac{2\eta}{mQ}.\ear
This implies
\bear\label{5.32}
\bar x(\eta,-{\textstyle\frac{1}{2}}\bar r)&>&l_4
(-{\textstyle\frac{1}{2}}\bar r)\qquad{\rm
for}
\ \ Q<2m,\nonumber\\
\bar x(\eta,-{\textstyle\frac{1}{2}}\bar r)&<&l_4
(-{\textstyle\frac{1}{2}}\bar r)\qquad{\rm
for} \ \ Q>2m.\ear
Thus, for $\eta\gg Q\gg m$, the $x^0$ integration in (\ref{3.25}) is
effectively cut off at $x^0=\bar x(\eta,-\bar r/2)$ and the $x^0$
integral in (\ref{3.25}) will get its leading contribution from the
scaling regime of $\widetilde{\cal M}_2(x^0,r)$ (\ref{5.30}). Inserting
(\ref{5.30}) in (\ref{3.25}) we get for $\eta\gg Q\gg m$
\bear\label{5.33}
T_2(i\eta,Q^2)&\sim&\frac{iMQ^2}{\eta}\int_C dr\int^\infty_0
dx^0(x^0)^2\nonumber\\
&&\times \exp(-x^0/\bar x(\eta,r))
A'(x^0M)^{a-1}(-r)^{-\frac{1}{2}\varepsilon_0}\nonumber\\
&\sim&i\frac{Q^2}{M^3}\left(\frac{M}{\eta}\right)^{3+a}\Gamma(2+a)
A' \int_C dr(-r)^{-\frac{1}{2}\varepsilon_0}(r+\bar
r)^{-2-a}\nonumber\\
&\sim&\frac{4\pi}{M}
A'\frac{\Gamma(1+a+\frac{1}{2}\varepsilon_0)}
{\Gamma(\frac{1}{2}\varepsilon_0)}
\left(\frac{Q^2}{2M^2}\right)^{-a-\frac{1}{2}\varepsilon_0}
\left(\frac{\eta}{M}\right)^{-1+a+\varepsilon_0}.\ear
The analytic continuation for arbitrary large $|\nu|$ is obtained by
the replacement in (\ref{5.33}):
\be\label{5.34}
\eta\to \nu\exp(-i{\textstyle\frac{1}{2}}\pi).\ee
For real positive $\nu$ this leads to
\bear\label{5.35}
\nu W_2(\nu,Q^2)&=& F_2(x_{Bj},Q^2)\nonumber\\
&\sim& 4\pi
A'\Gamma(1+a+{\textstyle\frac{1}{2}}\varepsilon_0)
(\Gamma({\textstyle\frac{1}{2}}\varepsilon_0))^{-1}
\nonumber\\
&&\times \sin\left[{\textstyle\frac{\pi}{2}}
(1-a-\varepsilon_0)\right]\left(\frac{Q^2}{2M^2}
\right)^{\frac{1}{2}\varepsilon_0}(x_{Bj})^{-a-\varepsilon_0}.\ear

This is an interesting result. The simple scaling assumption for the
matrix element $\widetilde{\cal M}_2$ (\ref{5.30}) as suggested by the
analogy to critical phenomena leads to the behaviour for the structure
function $F_2$ found in the Regge fit (see (2) and (4a) of
\cite{317}) for large $Q^2$ and small $x_{Bj}$:
\be\label{5.36}
F_2(x_{Bj},Q^2)\sim
X_0(Q^2_0)^{1+\varepsilon_0}\left(\frac{Q^2}{Q^2_0}\right)^{\frac{1}{2}
\varepsilon_0}(x_{Bj})^{-\varepsilon_0}.\ee
Here $X_0,Q^2_0,\varepsilon_0$ are constants with
\be\label{5.37}
\varepsilon_0\approx 0.44,\ee 
where $1+\varepsilon_0$ is the intercept of the hard pomeron. Setting
$a=0$ in (\ref{5.35}) we get indeed the powers of $Q^2$ and $x_{Bj}$
of (\ref{5.36}). Comparing (\ref{5.27}) and (\ref{5.29}) we see that
$\frac{1}{2}\varepsilon_0$  can be interpreted as the anomalous part
of a critical index in the language of statistical physics.

\section{Conclusions}
\setcounter{equation}{0}

In this article we have developed an approach for the
theoretical description of the structure functions at small $x_{Bj}$
which should allow truely nonperturbative calculations for all
$Q^2>0$. Of course, one can also perform perturbative calculations in
this framework.
We have introduced an effective, $r$-dependent theory (see sect. 4) in
 Minkowski and Euclidean space, starting from QCD in
Feynman-'t Hooft gauge. We have argued that the small $x_{Bj}$ behaviour of
the structure functions is related to the small $r$ behaviour of the
effective theory and that the limit $r\to 0$ corresponds to a critical
point. In the vicinity of this critical point we can expect to see
power behaviour of the relevant matrix elements $\widetilde{\cal
M}_{1,2}(x^0,r)$ (\ref{3.18}) with certain critical indices. We have
shown that  this leads to Regge behaviour of the structure functions
at small $x_{Bj}$ and large $Q^2$, as observed in the phenomenological
fits of \cite{317}. In this way the intercept of the hard pomeron is
related to a critical index of the $r$-dependent theory. We emphasize
that in principle both the hard and the soft pomeron contributions to
the structure functions should be calculable in our approach.

The idea that the small $x_{Bj}$ behaviour of the structure functions
may be related to some kind of critical behaviour has been suggested
by various authors. Our present article follows the ideas presented in
\cite{12}. The critical point of our effective theory for $r\to 0$, if
it is indeed confirmed, would be completely analogous to a point of a
second order phase transition in statistical physics.
Quite a different type of critical behaviour, self-organised
criticality,
was suggested for the small $x_{Bj}$ behaviour of the structure
functions in \cite{30a}. Criticality of the photon wave function in
connection with a dipole model  was suggested in \cite{312}. It is
also interesting to note that perturbative calculations in the leading
logarithmic approximation, that is in the framework of the BFKL
equation \cite{307}, lead to conformal invariant structures for the
amplitudes  \cite{30b,30c}. One can suspect that this conformal
invariance could have  its origin in some sort of critical behaviour
of an effective theory.

Coming back to our present article we have presented here also our
results on some more technical issues.

We have discussed various ways of
analytic continuation of the Compton amplitude in the $\nu$ and $r$
planes. We found the representations (\ref{3.19})-(\ref{3.25}) the
most useful ones. The $r$-dependent theory in Euclidean  space was
studied in particular in the Feynmann-'t Hooft gauge. The final goal
of our approach is to make a truely nonperturbative calculation of the
matrix elements (\ref{4.26}), (\ref{4.27}) in the $r$-dependent
Euclidean theory. This could be based for instance on exact
renormalization group methods (see \cite{20,21} for a review) or on
lattice methods. Certainly, it will not be an easy task, since our
Euclidean action (\ref{4.29}) has an imaginary part for $r\not=
1$. But suppose that one can indeed study in this way the theory first
for real $r$ with $0<r<2$, deduce critical behaviour 
for small $r$  and establish scaling relations
as in (\ref{5.29}). Making then the analytic continuation in $r$ and 
putting  everything into the  theoretical
machinery developed in this paper  would mean a calculation of the
small $x_{Bj}$ and large $Q^2$ behaviour of the structure
functions. Both the functional dependences and the absolute
normalisation of the structure functions could be obtained in this
way. Of course, much remains to be done.

\vspace{1.5cm}

\noindent {\bf Acknowledgements:} The author is grateful to 
M. Diehl, A. Donnachie,
H.G.~ Dosch,
C. Ewerz, A. Hebecker, P.V. Landshoff, E. Meggiolaro, 
C. Wetterich, and W. Wetzel for useful discussions and suggestions and 
to T. Paulus for technical assistance with the manuscript.

\newpage

\section*{Appendix A: Relation of $\widetilde{\cal M}^-_{a,b}$ and
$\widetilde{\cal M}^-_{1,2}$}
\def\theequation{A.\arabic{equation}}
\setcounter{equation}{0}
From the definitions (\ref{3.12b}) and (\ref{4.1}), (\ref{4.2})
and using (\ref{3.5}) we get after some straightforward but lengthy
algebra:
\bear\label{A.1}
\widetilde{\cal M}^-_a(x^0,r)&=&\bigl\{-3\frac{\partial^2}
{(\partial x^0)^2}
-\frac{6(1-r)}{x^0}\frac{\partial^2}
{\partial x^0\partial r}
\nonumber\\
&&+\frac{3r(2-r)}{(x^0)^2}\frac{\partial^2}
{(\partial r)^2}
-\frac{6r(2-r)}{(x^0)^2(1-r)}\frac{\partial}
{\partial r}\bigr\}
\widetilde {\cal M}^-_1(x^0,r)\nonumber\\
&&+\bigl\{-3M^2\frac{\partial^2}{(\partial x^0)^2}-
\frac{6(1-r)M^2}{x^0}\frac{\partial^2}
{\partial x^0\partial r}
+\frac{3r(2-r)-2}{(x^0)^2}M^2\frac{\partial^2}
{(\partial r)^2}\nonumber\\
&&-\frac{3r(2-r)-2}{(x^0)^2(1-r)}2M^2
\frac{\partial}{\partial r}
\bigr\}\widetilde{\cal M}^-_2(x^0,r),\ear
\be\label{A.2}
\widetilde{\cal M}_b^-(x^0,r)=-(x^0)^{-2}
\bigl\{\frac{\partial^2}{\partial r^2}-
\frac{2}{1-r}\frac{\partial}{\partial
r}\bigr\}
\{\widetilde{\cal M}_1^-(x^0,r)+M^2\widetilde{\cal M}_2^-(x^0,r)\}.
\ee

\section*{Appendix B: Effective Lagrangian density in covariant gauges}
\def\theequation{B.\arabic{equation}}
\setcounter{equation}{0}

Here we give the details of the derivation of ${\cal L}_r(x)$ (\ref{4.14}).
From the original Lagrangian ${\cal L}$ (\ref{4.9}) we get the canonical
momenta in the standard way. It is convenient to write first ${\cal L}$
separating time and space components of fields.
\bear\label{B.0}
{\cal L}&=&{\textstyle\frac{1}{2}}(\dot G^{aj}+\partial_jG^{a0}-gf_{abc}
G^{b0}G^{cj})\nonumber\\
&&(\dot G^{aj}+\partial_jG^{a0}-gf_{ab'c'}G^{b'0}G^{c'j})\nonumber\\
&&-{\textstyle\frac{1}{4}}G^a_{jk}G^{ajk}\nonumber\\
&&-{\textstyle\frac{1}{2\xi}}(\dot G^{a0}+\partial_jG^{aj})
(\dot G^{a0}+\partial_kG^{ak})\nonumber\\
&&+\sum_q\{\bar q{\textstyle\frac{i}{2}}\gamma^0\dot q-\dot{\bar q}{\textstyle\frac{i}{2}}\gamma^0q\nonumber\\
&&+\bar
q({\textstyle\frac{i}{2}}\gamma^j\stackrel{\leftrightarrow}{\partial}_j
-g\gamma^\mu G^a_\mu {\textstyle\frac{\lambda_a}{2}}-m_q)q\}\nonumber\\
&&+\dot{\bar\phi^a}\dot\phi^a-(\partial_j\bar\phi^a)\partial_j\phi^a\nonumber\\
&&-gf_{abc}\{\dot{\bar\phi^a}G^{b0}\phi^c+(\partial_j\bar
\phi^a)G^{bj}\phi^c\}.\ear
We get now easily

\be\label{B.1}
\Pi_{G^{a0}}=\frac{\partial {\cal L}}{\partial \dot G^{a0}}=-\frac{1}{\xi}
(\dot G^{a0}+\partial_jG^{aj}),\ee
\be\label{B.2}
\Pi_{G^{aj}}=\frac{\partial{\cal L}}{\partial\dot G^{aj}}=
\dot G^{aj}+
\partial_jG^{a0}-gf_{abc}G^{b0}G^{cj},\ee
\bear\label{B.3}
\Pi_q&=&\frac{\partial{\cal L}}{\partial\dot q}=
\bar q{\textstyle\frac{i}{2}}\gamma^0,\nonumber\\
\Pi_{\bar q}&=&\frac{\partial{\cal L}}{\partial\dot {\bar q}}=
-{\textstyle\frac{i}{2}}\gamma^0q,\ear
\bear\label{B.4}
\Pi_{\phi^a}&=&\frac{\partial{\cal L}}{\partial\dot {\phi^a}}
=\dot{\bar \phi^a},\nonumber\\
\Pi_{\bar{\phi^a}}&=&\frac{\partial{\cal L}}{\partial\dot{\bar
{\phi^a}}}
=\dot\phi^a-gf_{abc}G^{b0}\phi^c.\ear
Solving (\ref{B.1}), (\ref{B.2}), (\ref{B.4}) for the 
time derivatives of the fields, we get
\bear\label{B.5}
&&\dot G^{a0}=-\xi\Pi_{G^{a0}}-\partial_jG^{aj},\nonumber\\
&&\dot G^{aj}=\Pi_{G^{aj}}-\partial_jG^{a0}+gf_{abc}
G^{b0}G^{cj},\nonumber\\
&&\dot{\bar\phi^a}=\Pi_{\phi^a},\nonumber\\
&&\dot\phi^a=\Pi_{\bar\phi^a}+gf_{abc}G^{b0}\phi^c.\ear
The Hamiltonian density is
\bear\label{B.6}
{\cal H}&=&\Pi_{G^{a0}}\dot G^{a0}+\Pi_{G^{aj}}\dot G^{aj}
+\sum_q(\Pi_q\dot q+\dot{\bar q}\Pi_{\bar q})\nonumber\\
&&+\Pi_{\phi^a}\dot\phi^a+\dot{\bar\phi^a}\Pi_{\bar\phi^a}
-{\cal L},\ear
or written out explicitly:
\bear\label{B.7}
{\cal
H}&=&-{\textstyle\frac{1}{2}}\xi\Pi_{G^{a0}}\Pi_{G^{a0}}-\Pi_{G^{a0}}
\partial
_jG^{aj}\nonumber\\
&&+{\textstyle\frac{1}{2}}\Pi_{G^{aj}}\Pi_{G^{aj}}+
{\textstyle\frac{1}{4}}G^a_{jk}G^{ajk}\nonumber\\
&&+\Pi_{G^{aj}}(-\partial_jG^{a0}+gf_{abc}G^{b0}G^{cj})
\nonumber\\
&&+\sum_q\left\{\bar q(-{\textstyle \frac{i}{2}}\gamma^j
\stackrel{\leftrightarrow}{\partial}_j+g\gamma^\mu
G^a_\mu{\textstyle\frac{\lambda_a}{2}}+m_q)q\right\}\nonumber\\
&&+\Pi_{\phi^a}\Pi_{\bar\phi^a}+\Pi_{\phi^a}gf_{abc}G^{b0}\phi^c\nonumber\\
&&+(\partial_j\bar\phi^a)\partial_j\phi^a+gf_{abc}(\partial_j
\bar\phi^a)G^{bj}\phi^c.\ear
The $0-3$ component of the canonical energy-momentum tensor is
\bear\label{B.8}
{\cal T}_{03}&=&\Pi_{G^{a0}}\partial_3G^{a0}+\Pi_{G^{aj}}\partial_3
G^{aj}\nonumber\\
&&+\Pi_q\partial_3 q+(\partial_3\bar q)\Pi_{\bar q}\nonumber\\
&&+\Pi_{\phi^a}\partial_3\phi^a+(\partial_3\bar\phi^a)\Pi_{\bar\phi^a}.\ear
We have, furthermore,
\be\label{B.9}
P^3=\int d^3x{\cal T}^{03}=-\int d^3x{\cal T}_{03}\ee
and from the definitions (\ref{4.4}), (\ref{4.12})
\bear\label{B.10}
H_r&=&P^0-(1-r)P^3\nonumber\\
&=&\int d^3x\{{\cal H}+(1-r){\cal T}_{03}\}\nonumber\\
&=&\int d^3x{\cal H}_r.\ear
Thus, we get
\be\label{B.11}
{\cal H}_r={\cal H}+(1-r){\cal T}_{03}.\ee
Inserting here (\ref{B.7}) and (\ref{B.8}) leads to (\ref{4.13}).

To prove that ${\cal H}_r$ (\ref{4.13})
is the Hamiltonian density to the Lagrangian ${\cal L}_r(x)$ (\ref{4.14})
we proceed again in the standard way. We first write down ${\cal L}_r(x)$ 
separating time and space components of fields. 
\bear\label{B.12}
{\cal L}_r&=&
{\textstyle\frac{1}{2}}(\dot G^{aj}+\partial_jG^{a0}-gf_{abc}G^{b0}G^{cj}-(1-r)
\partial_3 G^{aj})\nonumber\\
&&(\dot G^{aj}+\partial_jG^{a0}-gf_{ab'c'}G^{b'0}G^{c'j}-(1-r)\partial_3
G^{aj})\nonumber\\
&&-{\textstyle\frac{1}{4}}G^a_{jk}G^{ajk}\nonumber\\
&&-{\textstyle\frac{1}{2\xi}}(\dot G^{a0}+\partial_jG^{aj}-(1-r)
\partial_3 G^{a0})
\nonumber\\
&&(\dot G^{a0}+\partial_kG^{ak}-(1-r)\partial_3G^{a0})\nonumber\\
&&+\sum_q\{\bar q{\textstyle\frac{i}{2}}\gamma^0\dot q-\dot{\bar q}
{\textstyle\frac{i}{2}}
\gamma^0q\nonumber\\
&&+\bar
q({\textstyle\frac{i}{2}}\gamma^j\stackrel{\leftrightarrow}{\partial}
_j-(1-r)
{\textstyle\frac{i}{2}}\gamma^0\stackrel{\leftrightarrow}{\partial}_3
-g\gamma^\mu G^a_\mu{\textstyle\frac{\lambda_a}{2}}-m_q)q\}\nonumber\\
&&+(\dot{\bar \phi^a} -(1-r)\partial _3\bar\phi^a)(\dot\phi^a-gf_{abc}G^{b0}
\phi^c-(1-r)\partial_3\phi^a)\nonumber\\
&&-(\partial_j\bar\phi^a)\partial_j\phi^a-gf_{abc}(\partial_j
\bar\phi^a)G^{bj}\phi^c.\ear
From (\ref{B.12}) we find the new canonical momenta as
\bear\label{B.13}
\Pi_{G^{a0}}&=&\frac{\partial{\cal L}_r}{\partial\dot G^{a0}}\nonumber\\
&=&-{\textstyle\frac{1}{\xi}}(\dot
G^{a0}+\partial_jG^{aj}-(1-r)\partial_3
G^{a0}),
\ear
\bear\label{B.14}
\Pi_{G^{aj}}&=&\frac{\partial{\cal L}_r}{\partial\dot G^{aj}}\nonumber\\
&=&\dot G^{aj}+\partial_jG^{a0}-gf_{abc}G^{b0}
G^{cj}-(1-r)\partial_3G^{aj},
\ear
\bear\label{B.16}
\Pi_{\phi^a}&=&\frac{\partial{\cal L}_r}{\partial\dot \phi^a}=\dot{\bar
\phi^a}-(1-r)\partial_3\bar\phi^a,\nonumber\\
\Pi_{\bar \phi^a}&=&\frac{\partial{\cal L}_r}{\partial\dot {\bar \phi^a}}
=\dot\phi^a-gf_{abc}G^{b0}\phi^c-(1-r)\partial_3\phi^a\ear
and $\Pi_q,\,\Pi_{\bar q}$ staying as in (\ref{B.3}).

Solving (\ref{B.13}), (\ref{B.14}), (\ref{B.16}) for the time
derivatives of the fields, we get:
\bear\label{B.17}
\dot G^{a0}&=&-\xi\Pi_{G^{a0}}-\partial_jG^{aj}+(1-r)\partial_3G^{a0},
\nonumber\\
\dot G^{aj}&=&\Pi_{G^{aj}}-\partial_jG^{a0}+
gf_{abc}G^{b0}G^{cj}+(1-r)\partial_3G^{aj},
\ear
\bear\label{B.18}
\dot{\bar\phi^a}&=&\Pi_{\phi^a}+(1-r)\partial_3\bar\phi^a,\nonumber\\
\dot\phi^a&=&\Pi_{\bar\phi^a}+gf_{abc}G^{b0}\phi^c+(1-r)\partial_3
\phi^a.\ear
With this we find that ${\cal H}_r$ (\ref{4.13})
is precisely given as
\bear\label{B.19}
{\cal H}_r&=&\Pi_{G^{a0}}\dot G^{a0}+\Pi_{G^{aj}}\dot G^{aj}
\nonumber\\
&&+\sum_q(\Pi_q\dot q+\dot{\bar q}\Pi_{\bar q})\nonumber\\
&&+\Pi_{\phi_a}\dot\phi^a+\dot{\bar\phi}^a\Pi_{\bar\phi^a}\nonumber\\
&&-{\cal L}_r.\ear

\section*{Appendix C: Effective Lagrangian density in the temporal gauge}
\def\theequation{C.\arabic{equation}}
\setcounter{equation}{0}

Here we give the $r$-dependent Lagrangian and Hamiltonian densities in
the temporal gauge.  We start from the standard Lagrangian of QCD and
add a total divergence term (see appendix E of \cite{14}).
\bear\label{C.1}
\tilde{\cal L} &=& -\frac{1}{4}  G^a_{\lambda\rho} G^{a
\lambda\rho}\nonumber\\
&& + \sum_q\bar q(i\gamma^\lambda D_\lambda-m_q)q\nonumber\\
&& + \partial_\lambda  \left\{\frac{1}{2}(G^{a \lambda }\partial_\rho
G^{a \rho }-G^{a \rho }\partial_\rho G^{a \lambda })
-\sum_q\bar q{\textstyle \frac{i}{2} }\gamma^\lambda q \right\},\nonumber\\
D_\lambda&=&\partial_\lambda+ig G^a_\lambda\frac{1}{2}\lambda_a.
\ear
With the gauge condition
\be\label{C.2}
G^a_0(x)=0,\quad (a=1,\ldots, 8),\ee
we get
\bear\label{C.3}
\tilde{\cal L}&=&\frac{1}{2}\dot G^{aj}\dot
G^{aj}-\frac{1}{2}(\partial_j
G^{ak})\partial_j G^{ak}+\frac{1}{2}(\partial_j G^{aj})\partial_k
G^{ak}\nonumber\\
&&-g\, f_{abc} G^{aj} G^{bk}\partial_j G^{ck}\nonumber\\
&&-\frac{1}{4} g^2 f_{abc} f_{ars} G^{bj}G^{ck} G^{rj}
G^{sk}\nonumber\\
&&+\sum_q\left\{\bar q\frac{i}{2}\gamma^0\dot q-\dot{\bar
q}\frac{i}{2}\gamma^0 q+\bar q\left(\frac{i}{2}\gamma^j
\stackrel{\leftrightarrow}{\partial}_j-m_q+g\gamma^j
G^{aj}\frac{\lambda_a}{2}\right)q\right\}.\ear
The canonical momenta are
\be\label{C.4}
\Pi_{G^{aj}}=\frac{\partial\tilde{\cal L}}{\partial\dot G^{aj}}=\dot
G^{aj}\ee
and $\Pi_q,\ \Pi_{\bar q}$ as in (\ref{B.3}). This leads to the
following Hamiltonian density and 0 - 3 component of the canonical
energy-momentum tensor:
\bear\label{C.5}
{\cal H}&=&\frac{1}{2}\Pi_{G^{aj}}\Pi_{G^{aj}}\nonumber\\
&&+\frac{1}{2}(\partial_k G^{aj})\partial_k G^{aj}\nonumber\\
&&-\frac{1}{2}(\partial_j G^{aj})\partial_k G^{ak}\nonumber\\
&&+g\, f_{abc} G^{aj} G^{bk}\partial_j G^{ck}\nonumber\\
&&+\frac{1}{4}g^2 f_{abc} f_{ast} G^{bj} G^{ck}G^{sj}G^{tk}\nonumber\\
&&+\sum_q\left\{\bar
q\left(-\frac{i}{2}\gamma^j\stackrel{\leftrightarrow}{\partial}_j
+m_q-g\gamma^j
G^{aj}\frac{\lambda_a}{2}\right)q\right\},\\
\label{C.6}
{\cal T}_{03}&=&\Pi_{G^{aj}}\partial_3 G^{aj}+\sum_q
\left\{\bar q\frac{i}{2}\gamma^0
\stackrel{\leftrightarrow}{\partial}_3 q\right\}.\ear
With this we get for the $r$-dependent Hamiltonian density
\bear\label{C.7}
{\cal H}_r&=&{\cal H}+(1-r){\cal T}_{03}\nonumber\\
&=&{\cal H}^{(0)}_r+{\cal H}^{{\rm Int}},\\
\label{C.8}
{\cal
H}_r^{(0)}&=&\frac{1}{2}\Pi_{G^{aj}}\Pi_{G^{aj}}+(1-r)\Pi_{G^{aj}}\partial_3
G^{aj}\nonumber\\
&&+\frac{1}{2}(\partial_j G^{ak})\partial_j G^{ak}\nonumber\\
&&-\frac{1}{2}(\partial_j G^{aj})\partial_k G^{ak}\nonumber\\
&&+\sum_q\left\{\bar
q\left(-\frac{i}{2}\gamma^j\stackrel{\leftrightarrow}
{\partial}_j+(1-r)\frac{i}{2}\gamma^0\stackrel{\leftrightarrow}{\partial}_3
+m_q\right)q\right\},\\
\label{C.9}
{\cal H}^{{\rm Int}}&=& g\, f_{abc} G^{aj} G^{bk}\partial_j
G^{ck}\nonumber\\
&&+\frac{1}{4} g^2f_{abc}f_{ast}G^{bj}G^{ck}G^{sj}G^{tk}\nonumber\\
&&-\sum_q\left\{\bar q g\gamma^j G^{aj}\frac{\lambda_a}{2} q\right\}.\ear
In the standard way we find from (\ref{C.8}), (\ref{C.9}) the
$r$-dependent Lagrangian density
\bear\label{C.10}
{\cal L}_r&=&{\cal L}^{(0)}_r+{\cal L}^{{\rm Int}},\\
\label{C.11}
{\cal L}_r^{(0)}&=&\frac{1}{2}(\dot G^{aj}-(1-r)\partial_3 G^{aj})
(\dot G^{aj}-(1-r)\partial_3 G^{aj})\nonumber\\
&&-\frac{1}{2}(\partial_k G^{aj})\partial_k G^{aj}\nonumber\\
&&+\frac{1}{2}(\partial_j G^{aj})\partial_k G^{ak}\nonumber\\
&&+\sum_q\left\{\bar q \left(\frac{i}{2}\gamma^\lambda
\stackrel{\leftrightarrow}{\partial}_\lambda-(1-r)\frac{i}{2}
\gamma^0\stackrel{\leftrightarrow}{\partial}_3-m_q\right)q\right\},\\
\label{C.12}
{\cal L}^{{\rm Int}}&=&-{\cal H}^{{\rm Int}}.
\ear
We also find with $\tilde{\cal L}$ from (\ref{C.3}).
\bear\label{C.13}
{\cal L}_r&=&\tilde{\cal L}-(1-r)\dot
G^{aj}\partial_3G^{aj}\nonumber\\
&&+\frac{1}{2}(1-r)^2(\partial_3 G^{aj})\partial_3 G^{aj}\nonumber\\
&&-(1-r)\sum_q\left\{\bar q\frac{i}{2}\gamma^0
\stackrel{\leftrightarrow}{\partial}_3q\right\}.
\ear

\section*{Appendix D: Euclidean Lagrangian density in covariant gauges}
\def\theequation{D.\arabic{equation}}
\setcounter{equation}{0}

Let $H_r$ be as in (\ref{4.12}), to be precise, we set $x^0=0$:
\be\label{D.1}
H_r=\int d^3 x{\cal H}_r(x)|_{x^0=0}.\ee
We define the Euclidean field operators with
\be\label{D.2}
V(X_4)=\exp(X_4H_r)\ee
as follows:
\bear\label{D.3}
G^a_{Ej}(X)&=&-V(X_4)G^{aj}(x)V(-X_4),\nonumber\\
G^a_{E4}(X)&=&-iV(X_4)G^{a0}(x)V(-X_4),\nonumber\\
q_E(X)&=&V(X_4)q(x)V(-X_4),\nonumber\\
\bar q_E(X)&=&V(X_4)\bar q(x)V(-X_4),\nonumber\\
\phi^a_E(X)&=&V(X_4)\phi^a(x)V(-X_4),\nonumber\\
\bar\phi^a_E(X)&=&V(X_4)\bar\phi^a(x)V(-X_4),\nonumber\\
\Pi_{G^a_{Ej}}(X)&=&-V(X_4)\Pi_{G^{aj}}(x)V(-X_4),\nonumber\\
\Pi_{G^a_{E4}}(X)&=& iV(X_4)\Pi_{G^{a0}}(x)V(-X_4),\nonumber\\
\Pi_{\phi^a_E}(X)&=&V(X_4)\Pi_{\phi^a}(x)V(-X_4),\nonumber\\
\Pi_{\bar\phi^a_E}(X)&=&V(X_4)\Pi_{\bar\phi^a}(x)V(-X_4).\ear
Here $G^{aj}(x)$ etc. on the r.h.s. of (\ref{D.3}) are the Minkowskian
field operators at the point
\be\label{D.4}
x=(\vec X,0)\ee
and $X=(\vec X,X_4)$.

Defining the Euclidean electromagnetic current operator as
\be\label{D.4a}
J_{E\mu}(X)=\sum_q\left\{Q_q\bar q_E(X)\gamma_{E\mu}q_E(X)\right\},\ee
where $Q_q$ are the quark charges, we get with $x$ as in  (\ref{D.4})
\bear\label{D.5}
J_{E4}(X)&=&V(X_4)J^0(x)V(-X_4),\nonumber\\
J_{Ej}(X)&=&-iV(X_4)J^j(x)V(-X_4).\ear
The nucleon field operators in the Euclidean theory are obtained from
(\ref{4.16}):
\bear\label{D.6}
A_{Es}(p,\tau)&=&V(\tau)A_s(p,0)V(-\tau),\nonumber\\
A^\dag_{Es}(p,\tau)&=&V(\tau)A^\dag_s(p,0)V(-\tau),\qquad
(s=\pm{\textstyle\frac{1}{2}}),
\ear
where we always take $p$ (\ref{2.14}) for a nucleon at rest. We have
\bear\label{D.7}
\lim_{\tau\to-\infty}A^\dag_{Es}(p,\tau)\mid0\rangle \exp(-M\tau)
& = & \mid N(p,s)\rangle,\nonumber\\
\lim_{\tau\to\infty}\langle 0\mid A_{Es}(p,\tau)\exp(M\tau)
& = & \langle N(p,s)\mid.
\ear

In the standard way we get now for the matrix elements $\widetilde
{\cal M}^-_{a,b}(-iX_4,r)$ defined in (\ref{4.7}),(\ref{4.8}) for $X_4>0$
\bear\label{D.8}
\widetilde{\cal M}_a^-(-iX_4,r) &=&
\frac{1}{2}\sum_s\langle N(p,s)\mid(-1)J_{E\mu}(\vec 0,X_4)J_{E\mu}(\vec
0,0)\mid N(p,s)\rangle \nonumber\\
&=& \frac{1}{2}\sum_s\lim_{{\tau_i\to-\infty}\atop{\tau_f\to+\infty}}
\exp[(\tau_f-\tau_i)M]\nonumber\\
&& \langle 0\mid A_s(p,\tau_f)(-1)J_{E\mu}(\vec 0,X_4)J_{E\mu}(\vec 0,0)
A_s^\dag(p,\tau_i)\mid 0\rangle .
\ear
\bear\label{D.9}
\widetilde{\cal M}_b^-(-iX_4,r) &=&
\frac{1}{2}\sum_s\langle N(p,s)\mid J_{E4}(\vec 0,X_4)J_{E4}(\vec
0,0)\mid N(p,s)\rangle \nonumber\\
&=& \frac{1}{2}\sum_s\lim_{{\tau_i\to-\infty}\atop{\tau_f\to+\infty}}
\exp[(\tau_f-\tau_i)M]\nonumber\\
&& \langle 0\mid A_s(p,\tau_f)J_{E4}(\vec 0,X_4)J_{E4}(\vec 0,0)
A_s^\dag(p,\tau_i)\mid 0\rangle .
\ear
The Euclidean Hamiltonian density is obtained from (\ref{4.13}) with
$x$ from (\ref{D.4})
\be\label{D.10}
{\cal H}_{Er}(X)=V(X_4){\cal H}_r(x)V(-X_4).
\ee
Using here (\ref{D.3}) leads to
\bear\label{D.11}
{\cal H}_{Er}(X) 
&=& \frac{1}{2}\xi\Pi_{G^a_{E4}}\Pi_{G^a_{E4}}+
\frac{1}{2}\Pi_{G^a_{Ej}}\Pi_{G^a_{Ej}}\nonumber\\
&&
-i\Pi_{G^a_{E4}}(\partial_jG^a_{Ej}+i(1-r)\partial_3G^a_{E4})\nonumber\\
&& +\Pi_{G^a_{Ej}}(i\partial_jG^a_{E4}+igf_{abc}G^b_{E4}G^c_{Ej}
   +(1-r)\partial_3G^a_{Ej})\nonumber\\
&& +\frac{1}{4}G^a_{Ejk}G^a_{Ejk}\nonumber\\
&& +\sum_q\{\bar q_E(\frac{1}{2}
\gamma_{Ej}\stackrel{\leftrightarrow}{\partial}_j+\frac{i}{2}\gamma_{E4}
(1-r)\stackrel{\leftrightarrow}{\partial}_3\nonumber\\
&& +ig\gamma_{E\mu}G^a_{E\mu}\frac{1}{2}\lambda_a+m_q)q_E\}\nonumber\\
&&
+\Pi_{\phi^a_E}\Pi_{\bar\phi^a_E}+\Pi_{\phi^a_E}
(igf_{abc}G^b_{E4}\phi^c_E+(1-r)\partial_3\phi^a_E)\nonumber\\
&&
+(1-r)(\partial_3\bar\phi^a_E)\Pi_{\bar\phi^a_E}+(\partial_j\bar\phi^a_E)
\partial_j\phi^a_E\nonumber\\
&& -gf_{abc}(\partial_j\bar\phi^a_E)G^b_{Ej}\phi^c_E,
\ear
where all fields on the r.h.s. are at point $X$ and we define the
Euclidean field strength tensor as
\be\label{D.12}
G^a_{E\mu\nu}=\partial_\mu G^a_{E\nu}-\partial_\nu G^a_{E\mu}-
gf_{abc}G^b_{E\mu}G^c_{E\nu}.
\ee
It is now straightforward to verify that ${\cal H}_{Er}$ is the
Euclidean Hamiltonian density to the Lagrangian density ${\cal
L}_{Er}$ (\ref{4.22}). Indeed, starting from (\ref{4.22}) we define
the Euclidean canonical momenta as follows, where $\dot
G^a_{E4}\equiv\partial_4G^a_{E4}$ etc.
\bear\label{D.13}
i\frac{\partial{\cal L}_{Er}}{\partial\dot G^a_{E4}} &=&
\Pi_{G^a_{E4}}\nonumber\\
&=& \frac{1}{\xi}\{i\dot G^a_{E4}+i\partial_jG^a_{Ej}-(1-r)\partial_3
    G^a_{E4}\},\nonumber\\
i\frac{\partial{\cal L}_{Er}}{\partial\dot G^a_{Ej}} &=&
\Pi_{G^a_{Ej}}\nonumber\\
&=& i\dot G^a_{Ej}-i\partial_j
G^a_{E4}-igf_{abc}G^b_{E4}G^c_{Ej}-(1-r)\partial_3
G^a_{Ej},\nonumber\\
i\frac{\partial{\cal L}_{Er}}{\partial\dot\phi^a_E} &=& 
\Pi_{\phi^a_E}\nonumber\\
&=& i\dot{\bar\phi^a_E}-(1-r)\partial_3\bar\phi^a_E,\nonumber\\
i\frac{\partial{\cal L}_{Er}}{\partial\dot{\bar\phi^a_E}} &=& 
\Pi_{\bar\phi^a_E}\nonumber\\
&=& i\dot\phi^a_E-igf_{abc}G^b_{E4}\phi^c_E
-(1-r)\partial_3\phi^a_E,\nonumber\\
i\frac{\partial{\cal L}_{Er}}{\partial\dot q_E} &=& 
\Pi_{q_E}=\bar q_E\frac{i}{2}\gamma_{E4},\nonumber\\
i\frac{\partial{\cal L}_{Er}}{\partial\dot{\bar q_E}} &=& 
\Pi_{\bar q_E}=-\frac{i}{2}\gamma_{E4}q_E.
\ear
The Hamiltonian density (\ref{D.11}) is now obtained from (\ref{4.22})
and (\ref{D.13}) as
\bear\label{D.14}
{\cal H}_{Er} &=& i\Pi_{G^a_{E\mu}}\dot G_{E\mu}\nonumber\\
&& +i\sum_q(\dot{\bar q}_E\Pi_{\bar q_E}+\Pi_{q_E}\dot q_E)\nonumber\\
&&
+i\Pi_{\phi^a_E}\dot\phi^a_E+i\dot{\bar\phi^a_E}\Pi_{\bar\phi^a_E}
+{\cal L}_{Er},
\ear
where the $X_4$-derivatives $\dot G_{E\mu}$ etc. have to be considered
as functions of the canonical momenta and the fields by inverting
(\ref{D.13}).

Having derived the connection of the Euclidean Hamilton and Lagrange
densities (\ref{D.11}) and (\ref{4.22}) we can use the standard
procedures of the path integral formalism to show that the matrix
elements (\ref{D.8}) and (\ref{D.9}) can be represented by the path
integrals (\ref{4.26}) and (\ref{4.27}).

\newpage
\begin{center}
\section*{Figures }
\end{center}
\begin{figure}[H]
  \unitlength1mm
  \begin{center}
    \begin{picture}(160,70)
      \put(10,0){\epsfig{file=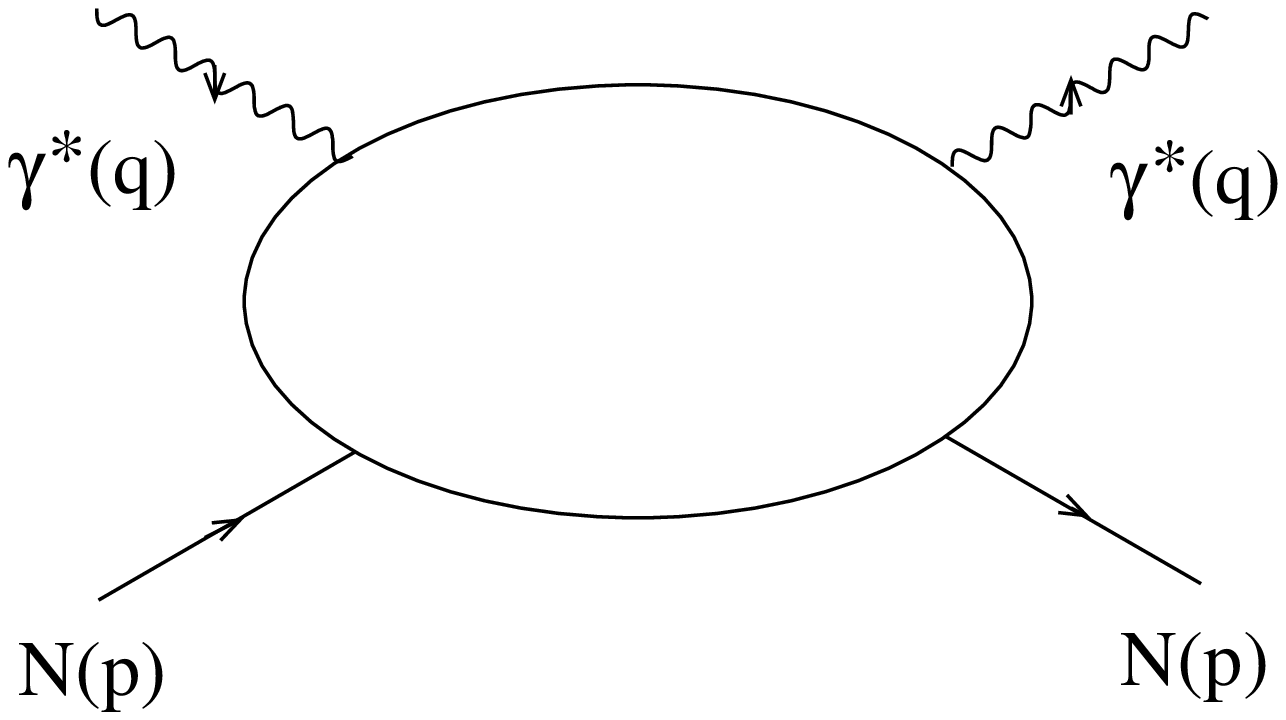,height=70mm}}
    \end{picture}
  \end{center}
  \caption{Virtual Compton scattering on a nucleon.}
\end{figure}

\begin{figure}[H]
  \unitlength1mm
  \begin{center}
    \begin{picture}(160,80)
      \put(20,0){\epsfig{file=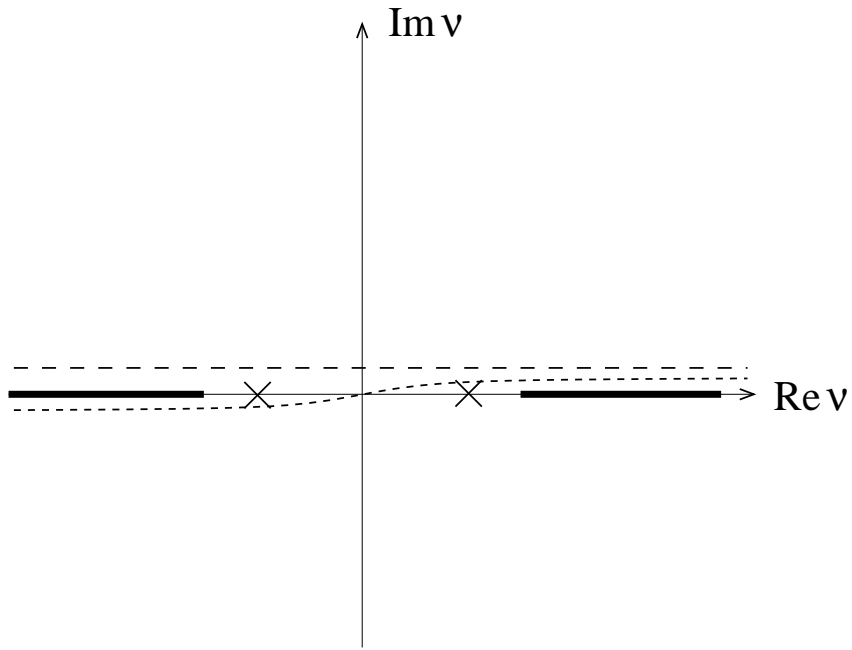,height=85mm}}
    \end{picture}
  \end{center}
  \caption{The $\nu$-plane with the position of the nucleon poles
$(\times)$ and the cuts. The funtions $T_j^{\rm F}(\nu,Q^2)\ (j=1,2)$ are
obtained along the short dashed line, $T_j^{{\rm ret}}(\nu,Q^2)$ along the
long dashed line.}
\end{figure}

\begin{figure}[H]
  \unitlength1mm
  \begin{center}
    \begin{picture}(160,150)
      \put(20,65){\epsfig{file=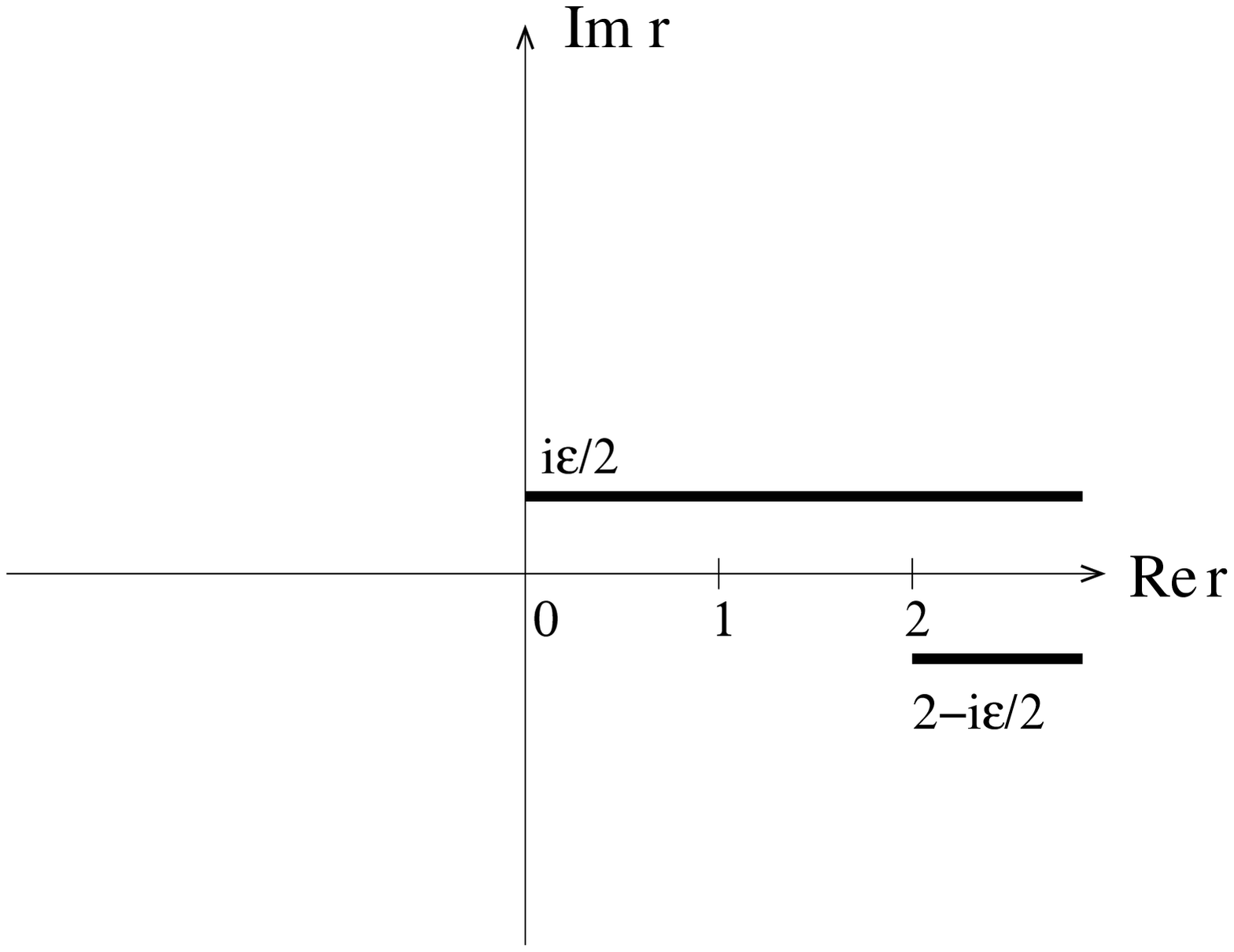,height=85mm}}
      \put(10,100){(a)}
      \put(20,-25){\epsfig{file=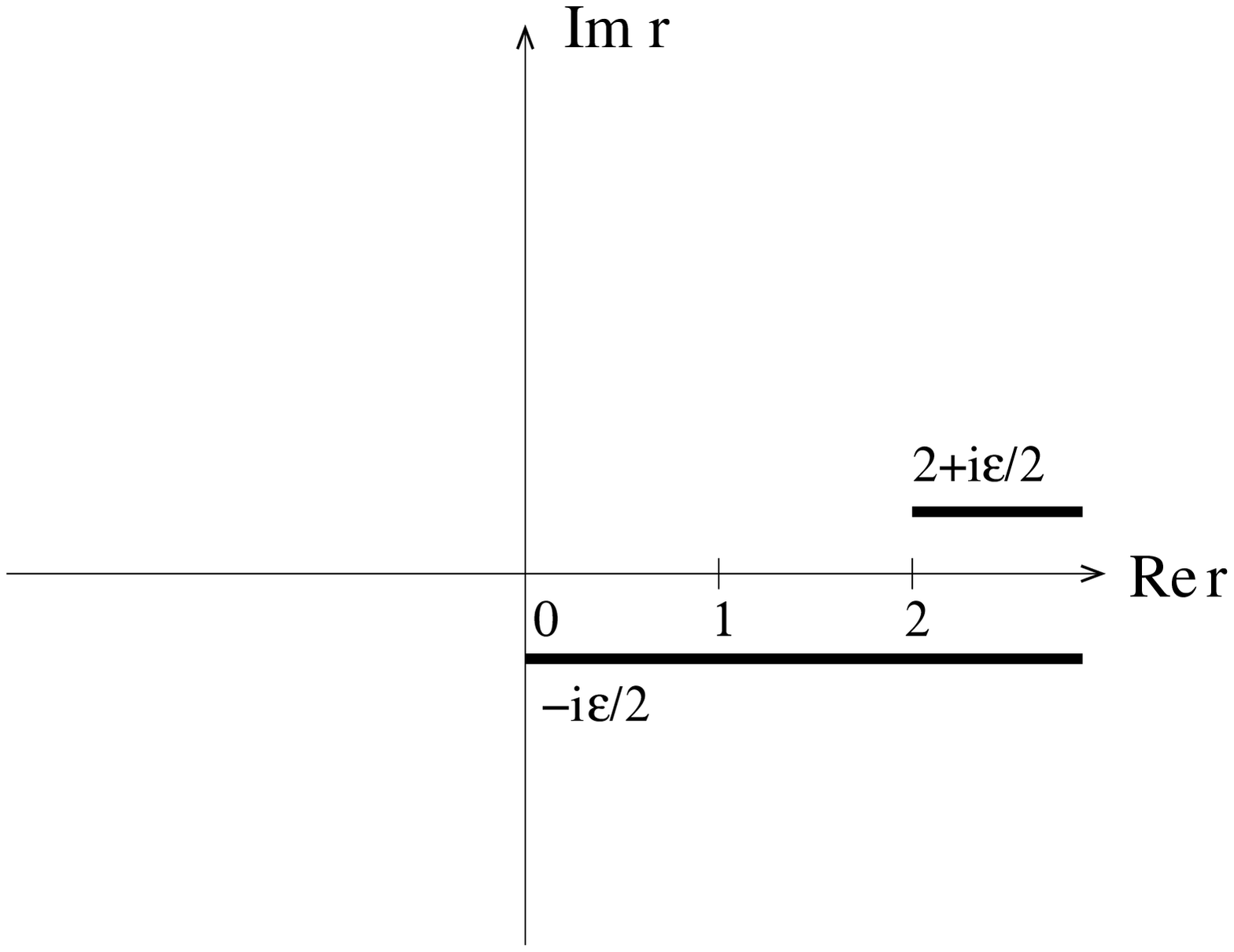,height=85mm}} 
      \put(10,10){(b)}
    \end{picture}
  \end{center}
  \vspace*{2.5cm}
  \caption{Cut structure of $\widetilde{\cal M}(x^0,r,\varepsilon)$ (a) and
$\widetilde{\cal M}^*(x^0,r^*,\varepsilon)$ (b) in the complex $r$-plane for
$0<\varepsilon<<1$.}
\end{figure}

\begin{figure}[H]
  \unitlength1mm
  \begin{center}
    \begin{picture}(160,85)
      \put(20,0){\epsfig{file=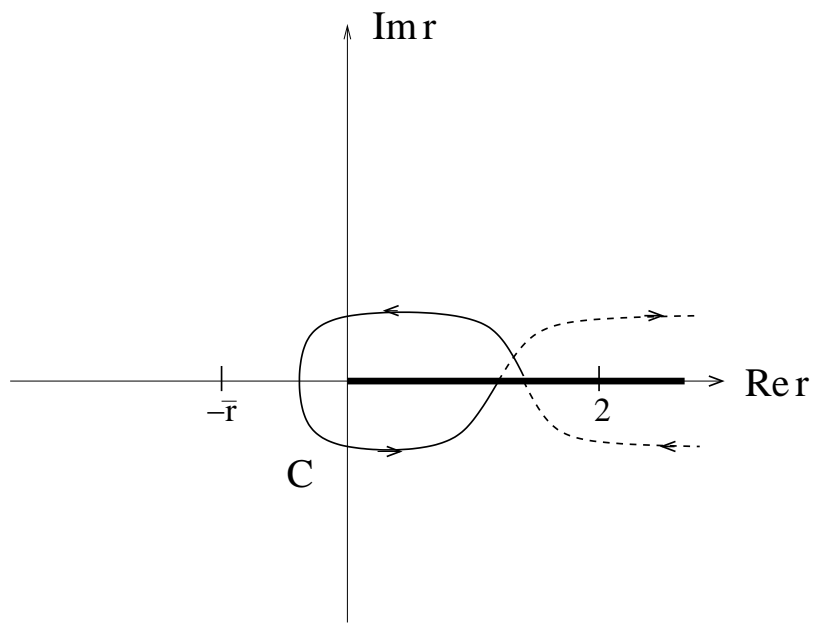,height=85mm}}
    \end{picture}
  \end{center}
  \caption{Cut structure of $\widetilde{\cal M}_j(x^0,r)$ and the curve $C$
in the complex $r$-plane. For $\bar r$ see (\ref{3.23}).}
\end{figure}

\end{document}